\documentclass[%
 reprint,
 amsmath,amssymb,
 aps,
prb,
twocolumn,
superscriptaddress
]{revtex4-2}
\usepackage{float} 
\usepackage{graphicx}
\usepackage{dcolumn}
\usepackage{tabularx}
\usepackage{amsmath}
\usepackage{bm}
\usepackage{hyperref}
\usepackage[mathlines]{lineno}

\newcommand{\gBCNO}{$\gamma$-${\rm Ba}_{3}{\rm CoNb}_{2}{\rm O}_{9}$}

\begin{document}

\preprint{APS/123-QED}

\title{Dynamical magnetism in the disordered cubic lattice material \gBCNO{}}



\author{Fanjun Xu}
 \affiliation{Helmholtz-Zentrum Berlin für Materialien und Energie GmbH, Hahn-Meitner Platz 1, D-14109 Berlin, Germany}%
 \affiliation{Institut für Festkörperphysik, Technische Universität Berlin, Hardenbergstraße 36, D-10623 Berlin, Germany}

\author{Ralf Feyerherm}
\affiliation{Helmholtz-Zentrum Berlin für Materialien und Energie GmbH, Hahn-Meitner Platz 1, D-14109 Berlin, Germany}

\author{Cecilie Glittum}
\affiliation{Department of Physics, University of Oslo, P.~O.~Box 1048 Blindern, N-0316 Oslo, Norway}
\affiliation{Helmholtz-Zentrum Berlin für Materialien und Energie GmbH, Hahn-Meitner Platz 1, D-14109 Berlin, Germany}
\affiliation{Dahlem Center for Complex Quantum Systems and Fachbereich Physik, Freie Universität Berlin, 14195 Berlin, Germany}

\author{Thomas J. Hicken}
\affiliation{PSI Center for Neutron and Muon Sciences CNM, 5232 Villigen PSI, Switzerland}

\author{Hubertus Luetkens}
\affiliation{PSI Center for Neutron and Muon Sciences CNM, 5232 Villigen PSI, Switzerland}

\author{Jonas A. Krieger}
\affiliation{PSI Center for Neutron and Muon Sciences CNM, 5232 Villigen PSI, Switzerland}

\author{Cintli Aguilar-Maldonado}
\affiliation{Helmholtz-Zentrum Berlin für Materialien und Energie GmbH, Hahn-Meitner Platz 1, D-14109 Berlin, Germany}

\author{Sven Luther}
\affiliation{Hochfeld-Magnetlabor Dresden (HLD-EMFL), Helmholtz-Zentrum
Dresden-Rossendorf (HZDR), Dresden 01328, Germany}

\author{Lucy K. Saunders}
\affiliation{Diamond Light Source, Harwell Science and Innovation Campus, Didcot, Oxfordshire, OX11 0DE, England, UK}

\author{Clemens Ritter}
\affiliation{Institut Laue-Langevin, 71 Av. des Martyrs, 38000 Grenoble, France}

\author{Peter Fouquet}
\affiliation{Institut Laue-Langevin, 71 Av. des Martyrs, 38000 Grenoble, France}

\author{Margarita Russina}
\affiliation{Helmholtz-Zentrum Berlin für Materialien und Energie GmbH, Hahn-Meitner Platz 1, D-14109 Berlin, Germany}

\author{Karel Prokeš}
\affiliation{Helmholtz-Zentrum Berlin für Materialien und Energie GmbH, Hahn-Meitner Platz 1, D-14109 Berlin, Germany}

\author{A.T.M. Nazmul Islam}
\affiliation{Helmholtz-Zentrum Berlin für Materialien und Energie GmbH, Hahn-Meitner Platz 1, D-14109 Berlin, Germany}

\author{Bella Lake}
\affiliation{Helmholtz-Zentrum Berlin für Materialien und Energie GmbH, Hahn-Meitner Platz 1, D-14109 Berlin, Germany}
\affiliation{Institut für Festkörperphysik, Technische Universität Berlin, Hardenbergstraße 36, D-10623 Berlin, Germany}

\date{\today}

\begin{abstract}

\gBCNO{} realizes a disordered simple-cubic spin-$1/2$ lattice in which Co$^{2+}$ ions randomly occupy one third of the sites, placing the system close to the site-percolation threshold for magnetic order. Specific-heat, susceptibility, neutron spin-echo, and muon spin-rotation measurements reveal a broad thermodynamic crossover, short-range magnetic correlations, and persistent fast spin dynamics down to at least 0.1~K, with no evidence for static order or conventional spin-glass freezing. Monte Carlo simulations yield a broad distribution of orphan spins, finite clusters, and an infinite network. The calculated orphan-spin fraction ($\approx 8.8\%$) agrees well with the weakly correlated spin fraction inferred from magnetization ($\approx 8.2\%$). Exact diagonalization of a diluted $S = 1/2$ Heisenberg model captures the broad magnetic specific-heat anomaly and supports the coexistence of weakly and strongly correlated spin environments. These results support a picture in which spin-$1/2$ quantum fluctuations, together with dilution and proximity to the percolation threshold, can support a disorder-driven dynamical state with short-range correlations in three dimensions, distinct from both classical spin glasses and geometrically frustrated quantum spin liquids.



\end{abstract}

\maketitle

\section{Introduction}

The search for unconventional magnetic ground states has largely focused on geometrically frustrated lattices, where competing interactions prevent ordering and can stabilize quantum spin liquids (QSLs). These are quantum many-body states experimentally characterized by the absence of long-range magnetic order and persistent spin fluctuations down to zero temperature~\cite{broholm2020quantum}. While QSL behavior is well established in one dimension and has been extensively explored in two dimensions~\cite{balz2016physical,shen2016evidence}, its realization in three-dimensional magnets remains both subtle and rare~\cite{chillal2020evidence, vzivkovic2021magnetic}. Crucially, the key experimental signatures of QSLs -- the absence of static order and persistent dynamics -- are not unique to geometrically frustrated systems: quenched disorder can, in principle, produce the same phenomenology through an entirely different mechanism \cite{savary2017disorder}. 

Over the past decades, it has become increasingly clear that the low-temperature magnetic behavior of spin-$1/2$ magnets can be profoundly modified by disorder. Contrary to the conventional expectation of a frozen spin-glass state, disordered antiferromagnets may instead exhibit strongly correlated spin fluctuations over an extended temperature range. Such disorder-stabilized dynamical states have been theoretically established in one dimension~\cite{fisher1994random} and extended to two dimensions~\cite{liu2020quantum}, where they can be distinguished from classical spin glasses by the absence of slow dynamics or static spin freezing. By contrast, theoretical and experimental studies of disorder-driven dynamics in three-dimensional quantum magnets remain scarce, and the interplay between dimensionality, quantum fluctuations, and quenched disorder is still poorly understood. 

The simple-cubic lattice provides a particularly transparent platform to address this issue. With nearest-neighbour interactions alone, this lattice is unfrustrated and, in the absence of disorder, is expected to develop conventional magnetic order. Partial site occupancy, however, introduces strong site disorder that can qualitatively alter the magnetic ground state. A physical realization of such a disordered cubic lattice is provided by the $\gamma$ polymorph of Ba$_3$CoNb$_2$O$_9$ (\gBCNO{}) \cite{treiber1982ordnungs}, in which magnetic Co$^{2+}$ ions randomly occupy only one third of the cube corners. This places the system just above the classical percolation threshold ($p_c \approx 0.31$ for the cubic lattice \cite{malarz2015simple}). In contrast to previously studied Cu-based analogues \cite{sana2024possible,hossain2024evidence} with quenched orbital moments, Co$^{2+}$ carries an effective spin-$1/2$ degree of freedom arising from the interplay of spin–orbit coupling and crystal-field effects, enabling studies of disorder-driven magnetic dynamics in systems of unquenched orbital moments.

In this article, we investigate the magnetic properties of \gBCNO{} using a combination of local and bulk experimental probes: specific-heat, magnetic susceptibility, neutron spin-echo (NSE), and muon spin-rotation ($\mu$SR) measurements. Experiments reveal a broad thermodynamic crossover, short-range spin correlations with a finite correlation length $\xi \approx 13$~\AA, and persistent fast spin dynamics down to at least 0.1~K, with no evidence for static order or conventional spin-glass freezing. We interpret this behavior within a connectivity-driven cluster framework in which isolated monomers, finite clusters, and one or more infinite network coexist with a broad distribution of local correlation strengths.

These findings identify \gBCNO{} as a three-dimensional diluted spin-$1/2$ magnet exhibiting short-range correlated dynamical magnetism near the percolation threshold. While such a state shares phenomenological features with quantum spin liquids, it emerges in a disordered simple cubic lattice without strong intrinsic geometric frustration, and therefore represents a distinct route to unconventional dynamical magnetism in three dimensions.



\section{Experiment}

Polycrystalline samples of \gBCNO{} were synthesized at the crystal lab of Core Lab Quantum Materials (CLQM), Helmholtz Zentrum Berlin für Materialien und Energie (HZB), Germany. Stoichiometric amounts of BaCO$_3$, CoO, and Nb$_2$O$_5$ were mixed and calcined in air at 1200~$^{\circ}$C for 12~h over 2-3 cycles, with intermediate pelletization and regrinding, followed by sintering at 1500$^{\circ}$C for 48~h in air. The sample was then cooled to room temperature at a rate of 400~$^{\circ}$C/h, yielding dark-brown powders. A single crystal was subsequently grown in an optical Floating Zone furnace (Crystal Systems Corp., FZ-T 10000-H-VI-VPO) with 500 W Tungsten halide lamps. It was performed in air at a growth rate of 4.0 mm/h. The as-grown single crystal was about 35 mm in length and about 5 mm in diameter. The \gBCNO{} sample used in all of the experiments was a fine powder carefully grounded from the single crystal of \gBCNO{}.

Phase purity was first verified by laboratory powder X-ray diffraction (Bruker D8, Cu K$\alpha_1$, $\lambda=1.54059$~\AA). To further confirm structural details, high-resolution synchrotron X-ray diffraction was performed at the I11 beamline of Diamond Light Source in the United Kingdom, using 15~keV photons ($\lambda=0.824037$~\AA) in transmission geometry with a MAC detector. Powders were diluted with boron–silica to reduce X-ray absorption and loaded into 0.3~mm boron–silica capillaries. Complementary neutron powder diffraction was carried out at the Institut Laue--Langevin (ILL), Grenoble, France, using the D2B ($\lambda = 1.59(4)$~\AA) and D20 diffractometers ($\lambda = 2.41$~\AA) at 1.5~K and room temperature on a 0.5~g powder sample sealed in a vanadium can.

Bulk thermodynamic and magnetic properties were characterized by specific heat and magnetization measurements. Heat capacity was measured from 300~K to 0.4~K in zero magnetic field using a Quantum Design PPMS equipped with a $^3$He insert (relaxation method with addenda subtraction). Magnetization was measured using a Quantum Design MPMS3 on powder samples. Zero-field-cooled and field-cooled curves were recorded at 1.8 and 350~K at $\mu_0H = 0.1$~T, and isothermal $M(H)$ curves were collected between –7 and 7~T for $T = 0.4$, 1.25 and 1.7~K.  
Additional high-field magnetization measurements in pulsed magnetic fields up to $\mu_0$H = 25 T and at $T= 0.67$~K were performed at the Hochfeld-Magnetlabor Dresden (HLD) with a compensated pick-up coil system in a pulsed-field magnetometer and a home-built $^{3}$He cryostat. Measurements were performed with the field applied along the [111] axes.
Normalization to absolute units was achieved by calibrating the pulsed-field data with the lower field PPMS magnetization obtained in static fields at the CLQM.

Magnetic correlations were investigated using NSE spectroscopy and $\mu$SR. NSE measurements were performed on the WASP spectrometer \cite{fouquet2007wide} at ILL at 1.5 and 20~K using an incident wavelength of approximately 4~\AA. The instrumental resolution was determined from elastic scattering of Ho$_{0.7}$Y$_{1.3}$Ti$_2$O$_7$. $\mu$SR experiments were carried out at the Paul Scherrer Institute (Villigen, Switzerland) using the FLAME instrument (0.1~K–10~K) and the GPS \cite{amato2017new} instrument (5–300~K). The initial muon-spin polarization was set in longitudinal geometry. Approximately 0.5~g of powder mixed with GE varnish was pressed into a pellet to improve thermal contact. Zero-field measurements were performed between 0.1~K and 300~K, and longitudinal-field measurements up to 3.2~T were carried out at 0.1~K.

\section{Results}

\subsection{Crystal structure}

\begin{figure}[]
    \centering
    \includegraphics[width=\linewidth]{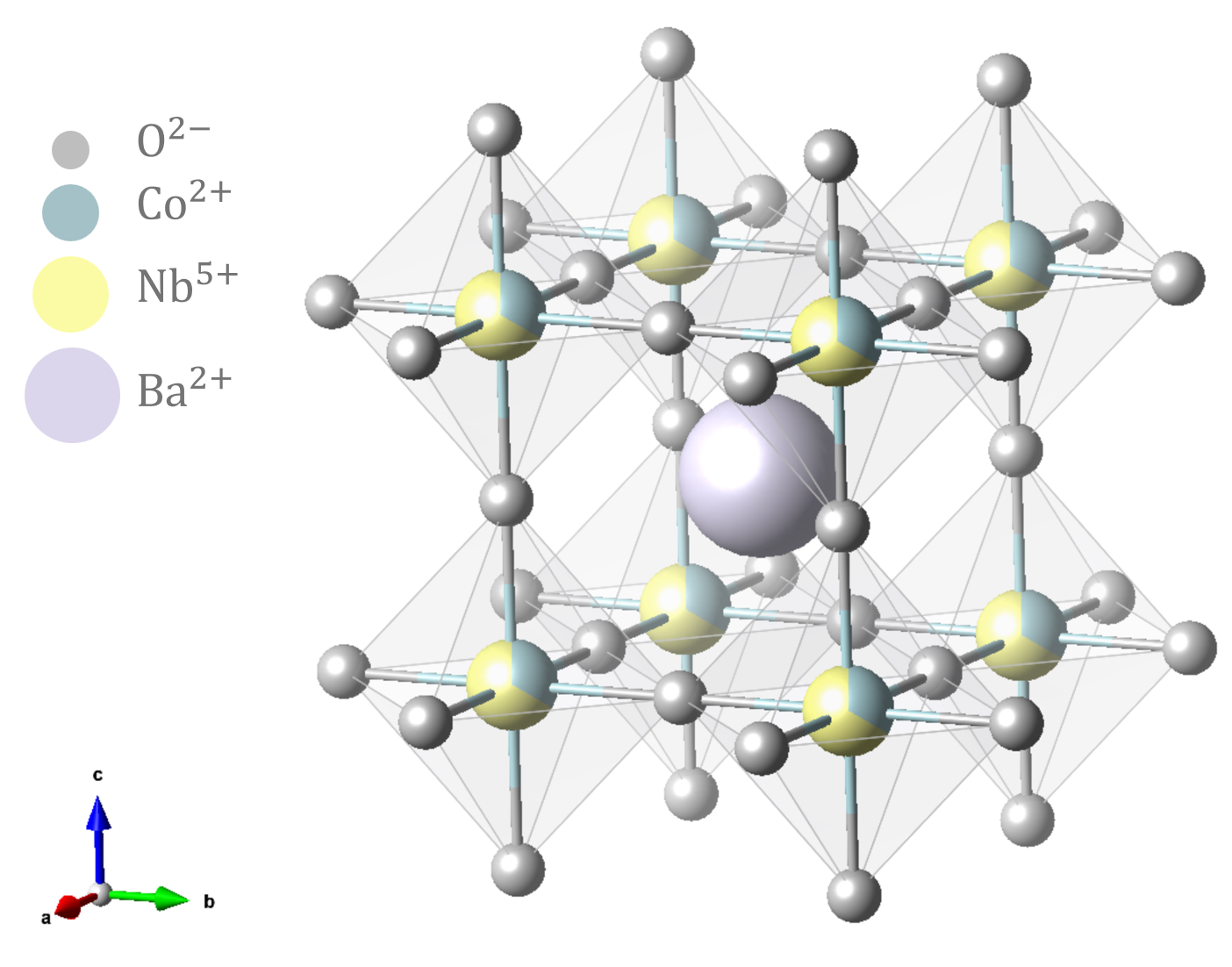}
    \caption{Atomic model of \gBCNO; space group Pm-3m.}
    \label{fig:Atomic model}
\end{figure}

Room temperature high resolution x-ray and neutron diffraction pattern (Appendix~\ref{sec:supplement}) show that \gBCNO{} has a cubic space group Pm$\bar{3}$m with $a = b = c = 4.0859(7)$ {\AA} at room temperature, the Ba${^{2+}}$ ions reside at the center (A-sites) of the cubic unit cell, and the Co${^{2+}}$/Nb${^{5+}}$ ions sit at the corners of the unit cell (B-sites) being surrounded by oxygen octahedra as shown in Fig.~\ref{fig:Atomic model}. The Co${^{2+}}$ and Nb${^{5+}}$ ions are 1:2 randomly distributed over the B site of the lattice. No structural phase transitions have been identified down to 1.5 K. 



\begin{figure}[]
    \centering
    \includegraphics[width=1\linewidth]{ 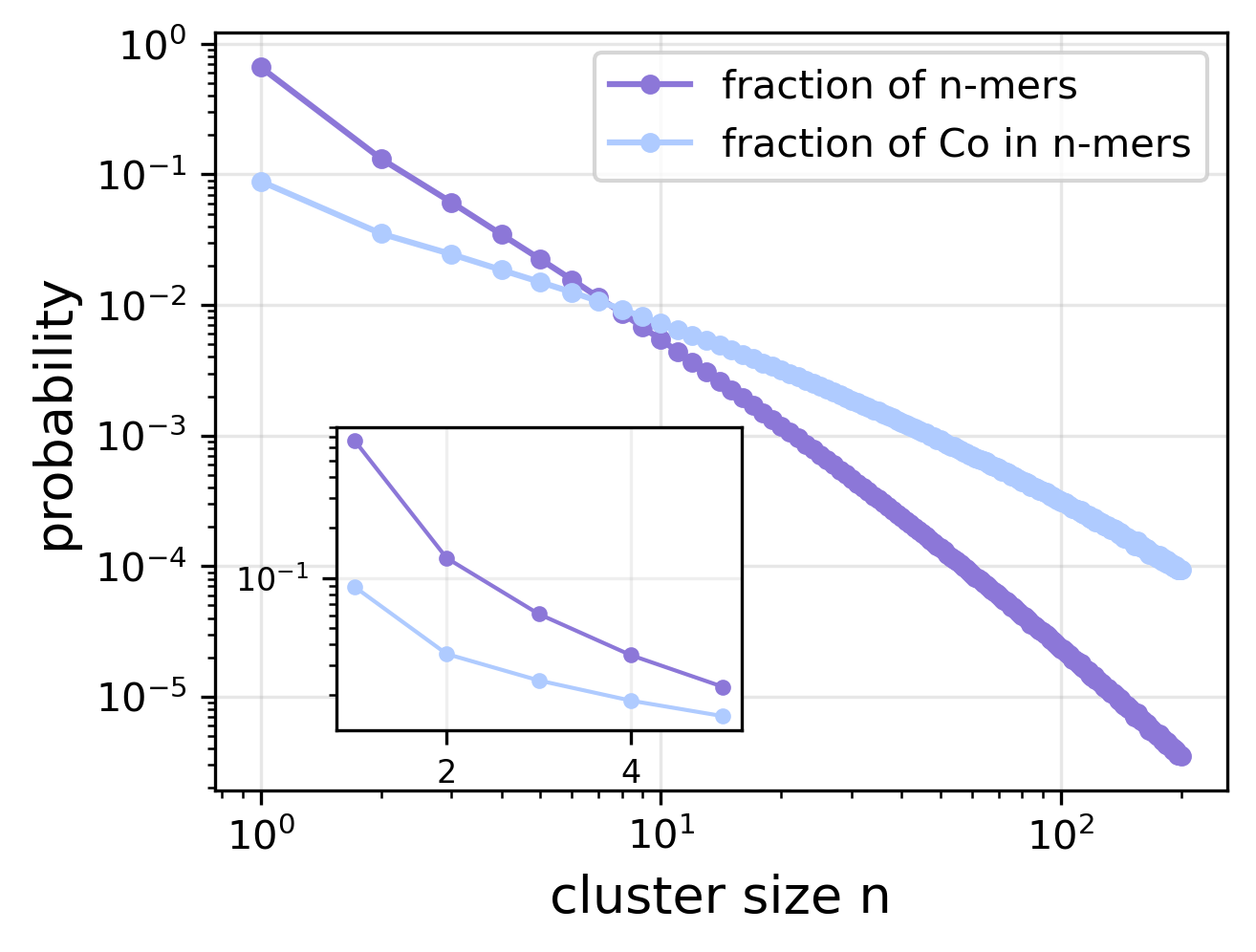}
    \caption{Monte Carlo simulation (cell size L$^{3}$=330$^{3}$) of  probabilities of clusters up to size n=200 (assuming first-neighbour interactions only), based on a $1/3$ diluted cubic lattice. The particle probability (blue) is the probability of a site being part of a cluster of size $n$. The relative frequency (purple) gives the probability of a cluster being of size $n$. The inset enlarges the small-$n$ regime ($n\le5$).}
    \label{fig:MC_cubic}
\end{figure}

To quantify the dilution effect on the cluster size in the cubic lattice statistically, we performed Monte Carlo simulations of a diluted cubic lattice with occupation probability $p=1/3$ (Fig.~\ref{fig:MC_cubic}). The results reveal a broad distribution of cluster-size based on nearest neighbours. Clusters with sizes $n\leq100$ collectively account for approximately 37\% of all Co sites, while at least one infinite well-connected network spanning the cell~\cite{stauffer2018introduction,aizenman1997number}, is always present, consistent with $p$ exceeding the cubic percolation threshold. The probability of monomer (orphan) spins is 8.78\%. These results establish the structural foundation for a magnetic lattice composed of orphan spins, finite clusters, and infinite networks.


\subsection{Thermodynamic properties}

\begin{figure}[]
    \centering
    \includegraphics[width=\linewidth]{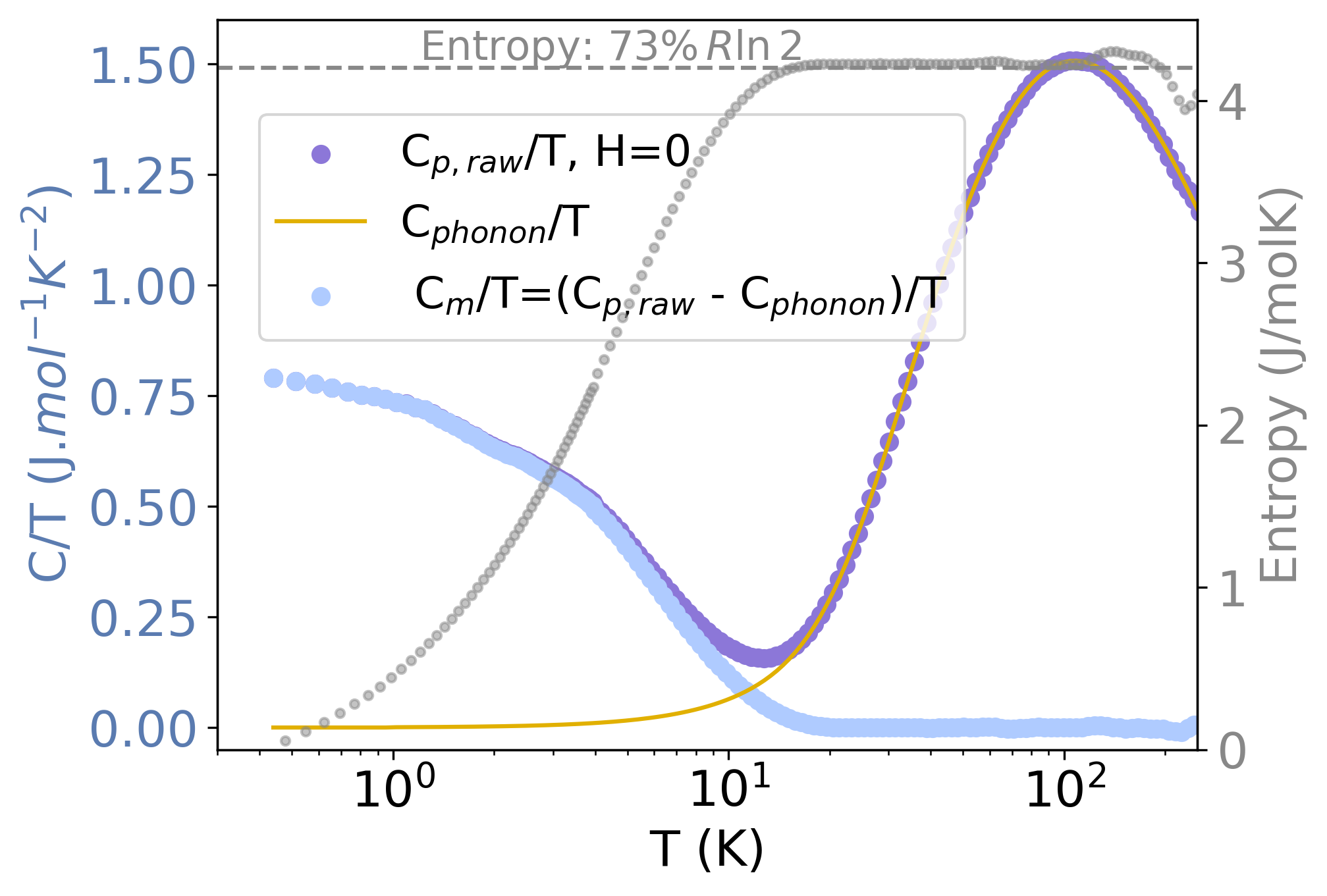}
    \includegraphics[width=\linewidth]{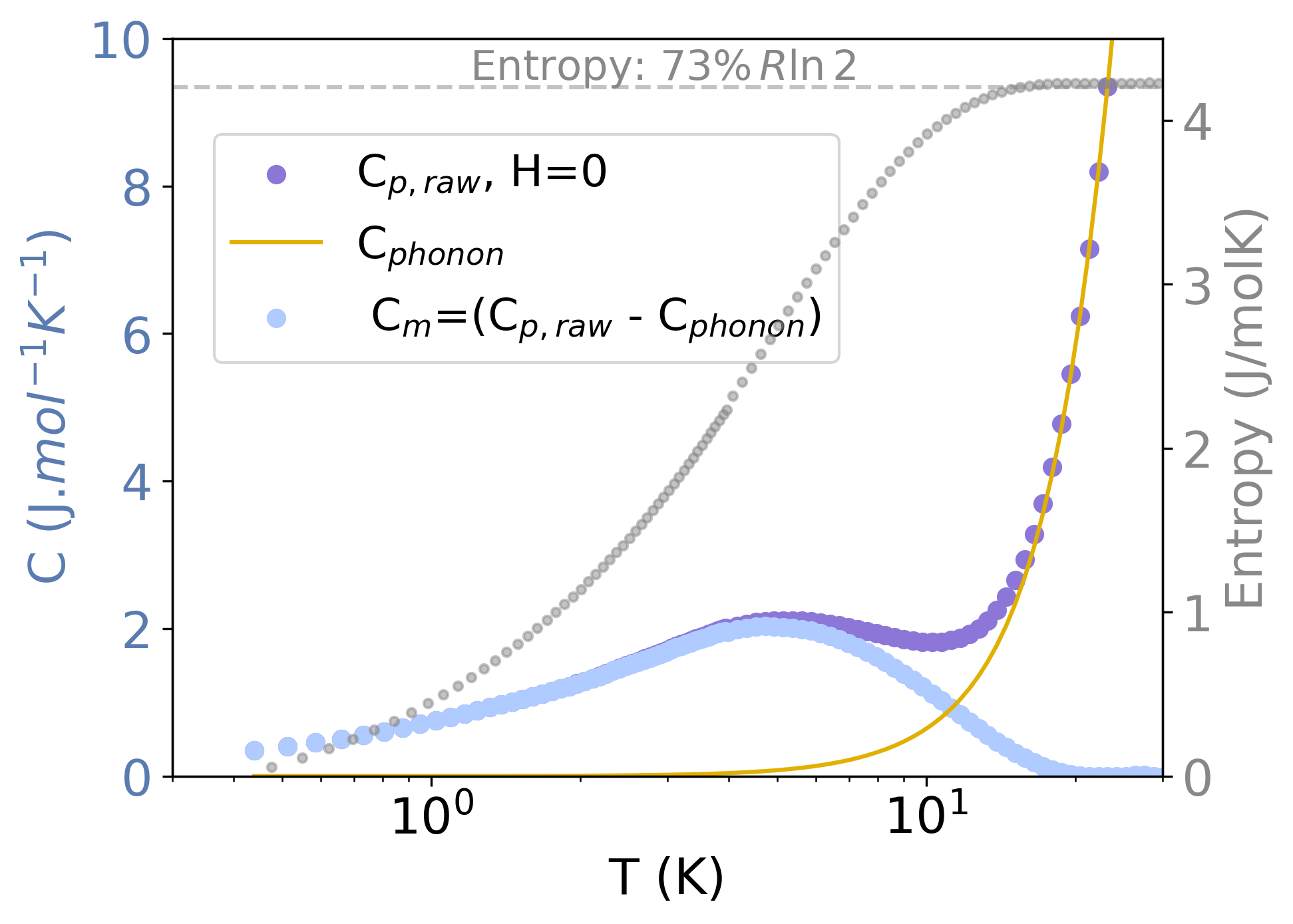}
    \caption{Temperature dependence of the zero-field heat capacity, showing the raw data, the fitted phonon contribution, the extracted magnetic contribution, and the corresponding magnetic entropy. (Top): $C(T)/T$.  (Bottom): $C(T)$.}
    \label{fig:Cp}
\end{figure}

Fig.~\ref{fig:Cp} shows the temperature dependence of the molar heat capacity in zero magnetic field. The absence of any sharp transition indicates the absence of long-range magnetic order down to 0.4 K. Instead, a broad maximum appears around 5 K, indicating the emergent short-range magnetic correlations at low temperature. The magnetic contribution is approximated by subtracting the phonon contribution from the total heat capacity. The phonon contribution is best approximated with one Debye and three Einstein modes:
\begin{align}
C_{p}&= 3 R N \Big[ (1 - \sum_{i=1}^{3} p_i)\, C_D(T, T_{D}) \\
  &\quad + \sum_{i=1}^{3} p_i \, C_E(T, T_{Ei}) \Big], \nonumber
\end{align}

with $C_{D}$ as the Debye term and $C_{E}$ as the Einstein term. By fitting the heat capacity between 15 K up to 257 K, the estimated phonon parameters are obtained: $T_{D}$ = 231 $\pm$ 22 K, $p_{1}$ = 0.354 $\pm$ 0.023, $T_{E1}$ = 362 $\pm$ 23 K, $p_{2}$ = 0.319 $\pm$ 0.025, $T_{E2}$ = 707 $\pm$ 18 K, $p_{3}$ = 0.057 $\pm$ 0.018, $T_{E3}$ = 147 $\pm$ 19 K.


The magnetic entropy is obtained by integrating the magnetic contribution $C_{m}/T$ from 0.4 K up to 15 K, 73\% of $R\ln2$ is recovered within this temperature range, consistent with an effective spin-$1/2$ system. The remaining entropy is of $27\%\,R\ln2$, indicates that not all magnetic degrees of freedom contribute fully to the entropy within this temperature window. This could reflect uncertainties in the phonon subtraction and/or a fraction of spins with weak couplings that contribute only partially to $C_{m}$ within this temperature window.




Fig. \ref{fig:MH-2terms} (Top) shows the dependence of the magnetization on the magnetic field up to 7 T at 0.4, 1.25 and 1.7 K. At 0.4~K, the low-field magnetization increases steeply and exhibits a change in slope around 2~T, above which it increases approximately linearly up to 7 T. However, the change around 2 T is quickly suppressed above 1 K. We tentatively attribute this 2 T feature to the presence of Co$^{2+}$ orphan spins. These are spins that are non- or weakly correlated and behave as nearly ideal paramagnetic moments. Within this framework, the 1.7 K magnetization measured at small applied fields can be described by the following expression: 



\begin{figure}[]
    \centering
    \includegraphics[width=1\linewidth]{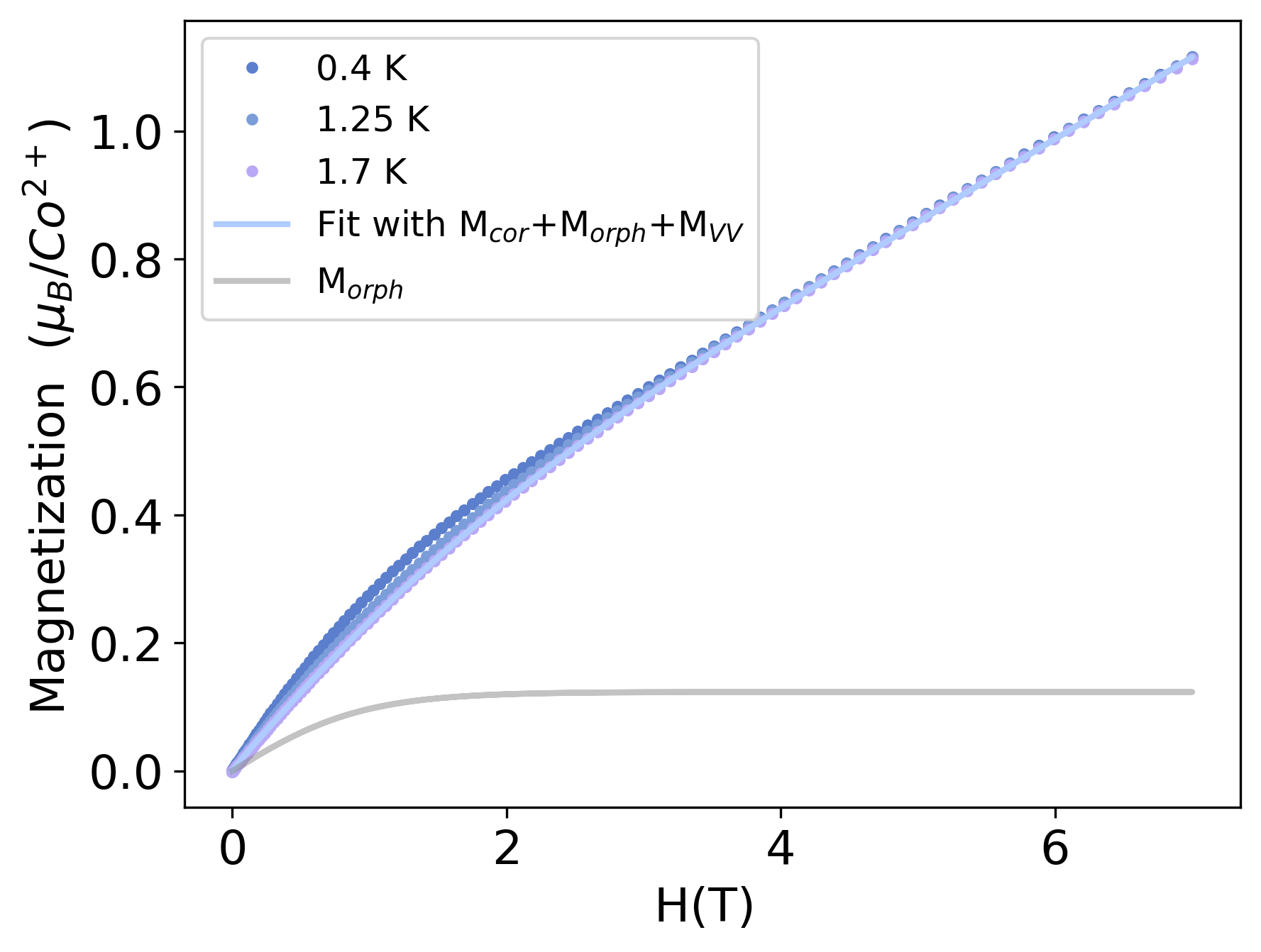}
    \includegraphics[width=1\linewidth]{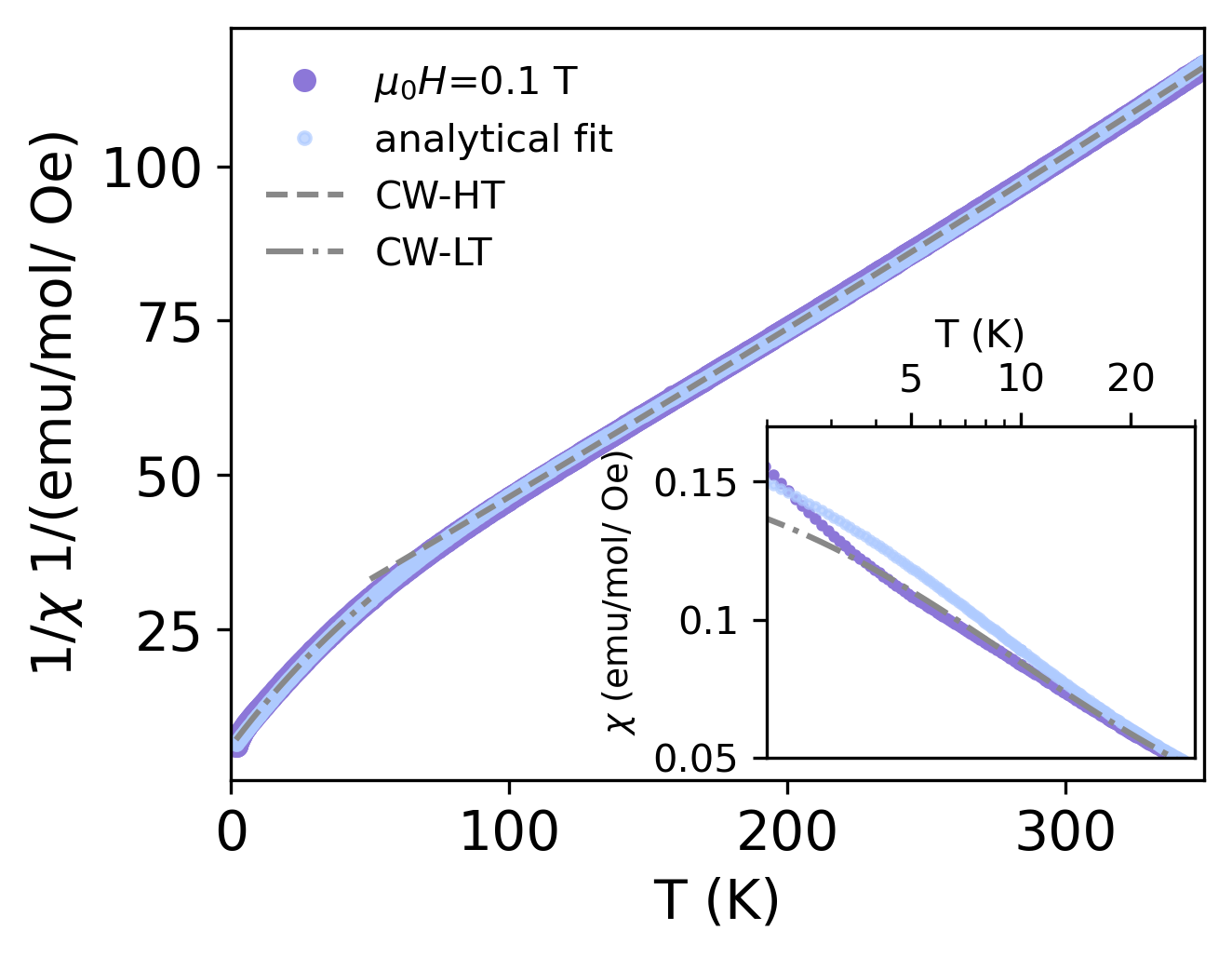}    
    \caption{Magnetization. (Top): Magnetization data measured at several temperatures, the data at $T = 1.7$ K is fitted with uncorrelated orphan spin and power law (Eq.\ref{MH}). (Bottom): Inverse susceptibility fit with different models as mentioned in the main text.}
    \label{fig:MH-2terms}
\end{figure}

\begin{equation}
\label{MH}
\begin{aligned}
M(H) =\ & \underbrace{p \cdot g\mu_{B}J B_{J}(x)}_{\text{Orphan\ spins}} \\
& +\ \underbrace{(1 - p) \cdot a H^b}_{\text{Correlated\ spins}} \\
& +\ \underbrace{\chi_{VV}\cdot H}_{\text{Van-Vleck\ term}}\hspace{-2.5mm},
\end{aligned}
\end{equation}
Here $H$ denotes the applied magnetic field. The first term, $B_J(x)$, is the modified Brillouin function describing the orphan-spin contribution, with $x = Jg\mu_{B}H/[k_{B}(T+\theta_{\rm Orphan})]$, for effective spin number $J=1/2$:
\begin{align}
B_{1/2}(x) = \tanh(x).
\end{align}
The second term describes the correlated spins ensemble (clusters + infinite spin network). The third term is a Van-Vleck paramagnetic contribution which is determined from the high field magnetization data as described in the Appendix~\ref{app:Highfield}. The magnetization analysis indicates a weakly correlated (effectively orphan-like) spin fraction p of around 8.2(2)\%. These moments exhibit a small net antiferromagnetic interaction, parameterized by $\theta_{\rm Orphan} \approx -0.86(2)~K$, together with an effective $g$ factor of 3.98(5) and power-law exponent $b = 0.92(9)$.



Fig.~\ref{fig:MH-2terms} (bottom) shows the inverse magnetic susceptibility from 2 K to 300 K. It exhibits approximately linear temperature dependence above 100 K, a crossover between 50 and 100 K, and a second approximately linear regime down to about 10 K, below which clear deviations appear. The crossover between 50 and 100 K is commonly observed in Co$^{2+}$--based compounds and reflects the interplay of spin-orbit coupling and crystal-electric-field effects in an octahedral oxygen environment.

As the first approximation, the two linear regions: high temperature region (from 100 K to 350 K) and intermediate temperature region (from 24.3 K to 66.5 K), are fitted with the Curie-Weiss law
\begin{align}
    \chi = \frac{C}{T - \theta_{\rm CW}},
\end{align}
with $C = N_{A}\mu_{0}\mu_{B}^{2}\mu_{\rm eff}^{2}/(3k_{B})$, and Curie-Weiss temperature $\theta_{cw}$ . This simplest model provides the first approximation to the data while ignoring all the microscopic details (spin--orbit coupling and crystal--electric field (SOC--CEF), crystal structure defect, etc). The fits give $\theta_{\mathrm{CW}}$ values of approximately $-10$ K in the intermediate regime and $-80$ K in the high-temperature regime. The high-temperature Curie--Weiss temperature is not physically meaningful as a direct measure of exchange interactions, but instead mainly reflects the influence of the SOC--CEF effect \cite{mugiraneza2022tutorial}. Consistently, the effective moment extracted at high temperatures is enhanced compared to its low-temperature value (see table~\ref{tab:susceptibility_parameter}), arising from the thermal population of higher-lying crystal-field levels beyond the ground-state doublet.  



To account for the effect of SOC--CEF more explicitly, we next modeled susceptibility using the analytical expression of \cite{lloret2008magnetic, lueken2013magnetochemie} over a wide temperature range:

\begin{align}
\chi_{\mathrm{SOC-CEF}}
= 
scale\cdot\frac{N_A\,\mu_0\,}{3 k_B (T-\theta_{\rm CW})}\,
\mu_{\mathrm{eff}}^{\,2}(T,\lambda),
\end{align}
  
where the temperature-dependent effective moment $\mu_{\mathrm{\rm eff}}(T,\lambda)$ arises from SOC--CEF level splitting and depends on the SOC--CEF coupling parameter $\lambda$. The analytical expression for $\mu_{\rm eff}$ is included in the Appendix~\ref{app:Comoment}. The fitted temperature range is 24.3 K to 350 K. This model yields reasonable agreement with the data and gives an effective Curie–Weiss temperature, $\theta_{\mathrm{CW}} \approx -10.6$ K.

In the fitting procedure, an overall scale factor for the magnetic moment was included, yielding a refined value of 0.947. The SOC--CEF coupling parameter was constrained to the physically allowed range $-246.8 \leq \lambda \leq 0$~K, where $\lambda = -246.8$~K corresponds to the isotropic limit of the Co$^{2+}$ ion~\cite{lueken2013magnetochemie}. Any local crystal-field distortion reduces the magnitude of the effective SOC parameter, leading to values $\lambda > -246.8$ K. The fit converges to this lower bound, $\lambda = -246.8$~K, indicating that deviations from the isotropic limit are not resolved within the present analysis.




In addition, the inset of Fig. \ref{fig:MH-2terms} shows a further low-temperature crossover. Even after including SOC--CEF effects, significant deviations remain below about 10 K, where magnetic correlations become stronger.

\begin{table}[]
    \caption{Fitting of magnetic susceptibility between 2 K to 300 K with different models.}
    \label{tab:susceptibility_parameter}
    \centering
    \begin{tabular}{ |c|c|c| } 
           \hline
         Model & $\theta_{\rm CW}$ (K) & $\mu_{\rm eff}$ ($\mu_{B}$) \\
         \hline
         \hline
          Curie Weiss fit (LT)& -9.68 $\pm$ 0.03 & 3.464 $\pm$ 0.002\\
          Curie Weiss fit (HT)& -77.5 $\pm$ 0.14 & 5.596 $\pm$ 0.003 \\
          Analytical SOC-CEF fit& -10.6 $\pm$ 0.11 & 3.753 $\pm$ 0.001\\
         \hline
         \end{tabular}
\end{table}

Across both fitting approaches, the extracted Curie–Weiss temperature is of the order $\sim -10$~K, indicating dominant antiferromagnetic interactions. The improved performance of SOC--CEF–based models over a simple Curie–Weiss description highlights the crucial role of spin–orbit coupling and crystal-field excitations in determining the magnetic response of Co$^{2+}$.

In the real material, random local distortions of the oxygen octahedra inevitably lead to spatial variations in SOC--CEF splitting parameters. As a result, the extracted parameters should be regarded as effective, average quantities rather than precise microscopic values. Nevertheless, both the modified two-level model and the analytical SOC–CEF expression provide a physically consistent and quantitatively reasonable description of the susceptibility data.

\subsection{Magnetic neutron diffraction and neutron spin echo}

\begin{figure}[]
    \centering
    \includegraphics[width=\linewidth]{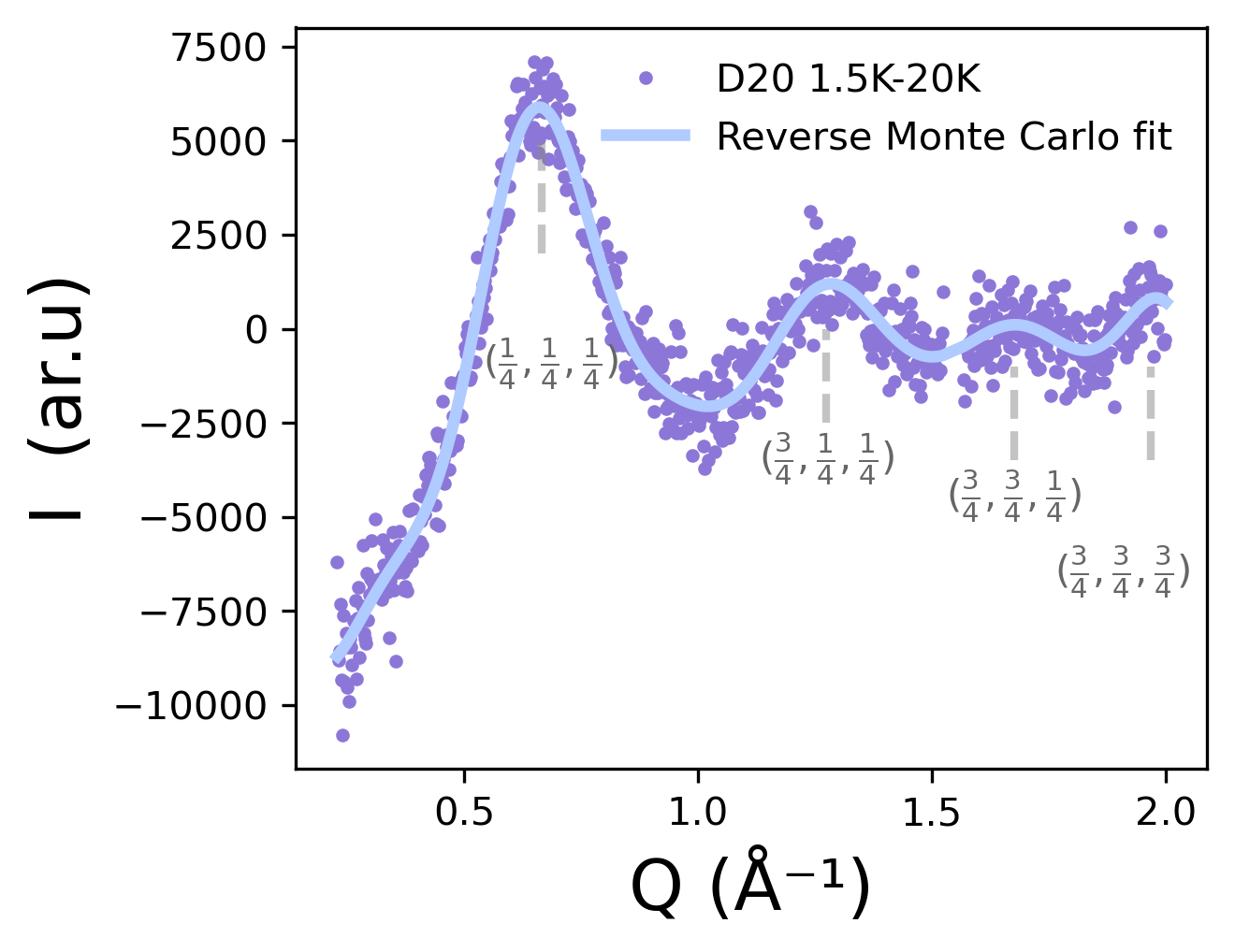}
    \includegraphics[width=\linewidth]{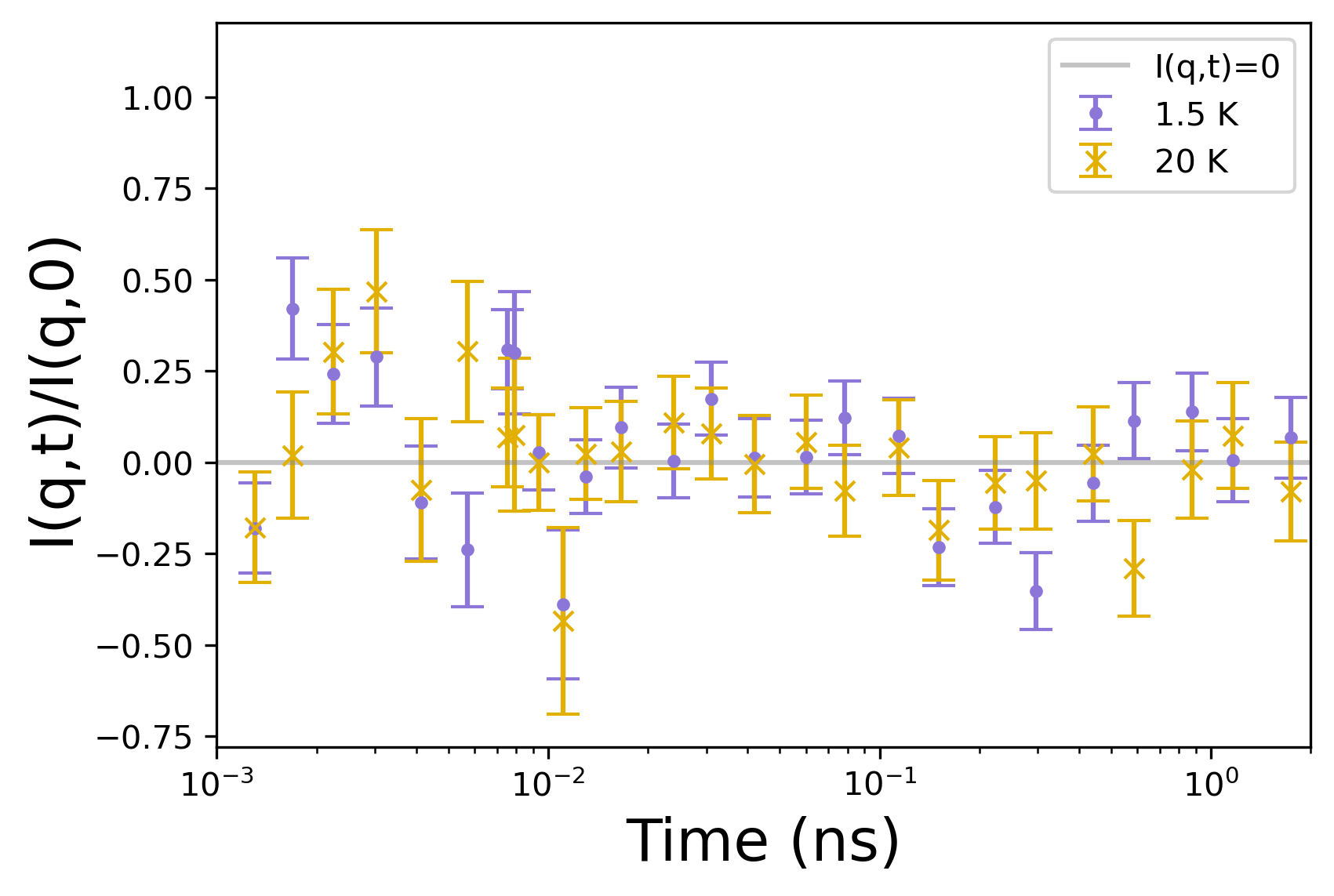}    
    \caption{ (Top): Magnetic diffuse scattering and reverse Monte Carlo refinement obtained using \textsc{Spinvert}~\cite{paddison2013spinvert}, with a simulation box of size $15\times15\times15$. The refinement was repeated 15 times and the resulting spin configurations were averaged. (Bottom): NSE intermediate scattering function at $Q = 0.64 \,\mathrm{\AA^{-1}}$. A finite magnetic signal is detected by polarization analysis, as indicated by the measurable polarization and the finite statistical uncertainties of the spectra. Within the accessible time window (1~ps--1~ns) the echo polarization is already fully relaxed, and no time-independent component with finite polarization is resolved within the experimental precision.} 
    \label{fig:spinvert results}
\end{figure}

By subtracting the 20~K neutron diffraction data from the low-temperature data, the magnetic scattering contribution shown in Fig.~\ref{fig:spinvert results} is obtained. 

At higher temperatures, there is featureless paramagnetic scattering with a $Q$-dependence reflecting the Co$^{2+}$ magnetic form factor. Upon cooling, the development of short-range correlations redistributes the spectral weight into broad diffuse peaks at well-defined momentum transfers. At low temperatures, the dominant diffuse maximum is centered at $Q \approx 0.67$~\AA$^{-1}$, and the first four broad features can be indexed by wave vectors of the form $(h\pm\frac{1}{4},k\pm\frac{1}{4},l\pm\frac{1}{4})$, consistent with short-range correlations characterized by a propagation vector ${\bf k}=(\frac{1}{4},\frac{1}{4},\frac{1}{4})$.

The real-space spin--spin correlation function extracted from the diffraction data decays over a finite length scale (Fig.~\ref{fig:SSresult}), yielding a correlation length of $\xi \approx 13.3$~\AA, i.e. only a few unit cells. This confirms that the magnetic correlations remain short-ranged down to 1.5 K.



Such short-range correlations are observed in a wide range of disordered or strongly correlated magnetic systems, including spin glasses, spin ices, and spin liquids. Their dynamical responses, however, differ qualitatively. Spin glasses and spin ices typically exhibit slow spin dynamics, leading either to a clear relaxation of the spin--spin correlation function or to a static component within the time window of conventional NSE~\cite{pickup2009generalized,ehlers2006study}. These slow dynamics can remain visible even at temperatures exceeding twice the Curie--Weiss temperature. Spin liquids, by contrast, are generally characterized by the absence of a static spin component. Establishing the relevant fluctuation time scale is therefore essential for distinguishing among these states.

To assess the dynamical character of these short-range correlations, NSE measurements were performed at $Q=0.64$~\AA$^{-1}$, close to the primary diffuse maximum. The normalized intermediate scattering function $I(Q,t)$ obtained from the polarization analysis is shown in the bottom panel of Fig.~\ref{fig:spinvert results}.


Within the accessible time window of the experiment ($1~\mathrm{ps}$ to $1~\mathrm{ns}$), $I(Q,t)$ remains close to zero with no systematic time dependence. This behavior is observed both below and above $\theta_{\rm CW}$, with no discernible temperature dependence within the experimental uncertainty. The data therefore indicate that the dominant spin relaxation processes occur on timescales shorter than the NSE window probed here, implying an upper bound of $\tau \lesssim 1$~ps for the relevant correlation time at the diffuse peak wavevector.

Importantly, no time-independent component with finite polarization is detected within the experimental precision. This absence of a static contribution contrasts with canonical spin-glass or spin-ice systems, in which slow dynamics or frozen moments produce a finite plateau in $I(Q,t)$ within the NSE time window~\cite{pickup2009generalized,ehlers2006study,FJXUNSE2026}. Instead, the results suggest that the short-range correlations revealed by neutron diffraction remain dynamic on the time scales accessible to NSE.





\subsection{Muon Spin Rotation}

$\mu$SR has a slower but partially overlapping time window compared with neutron diffraction and NSE, providing a complementary sensitivity to the low-energy magnetic dynamics. To further elucidate the nature of the above-observed feature (short-range, fast fluctuating spins rather than glassy behavior), Zero-field (ZF) and Longitudinal fields (LF) measurements were carried out.

\subsubsection{Zero-Field}

\begin{figure}[]
    \centering
    \includegraphics[width=1\linewidth]{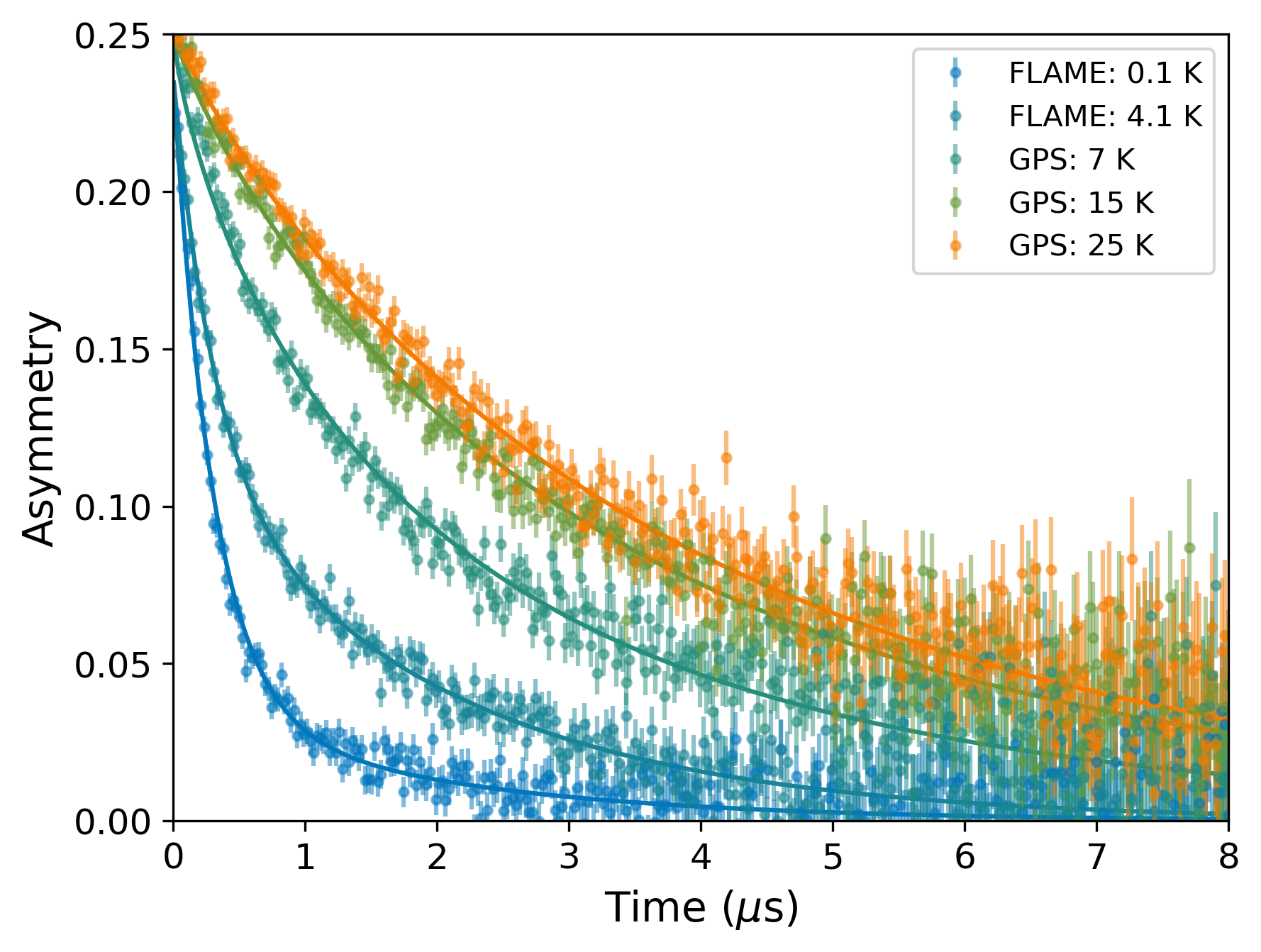}
    \vspace{2mm}
    \includegraphics[width=1\linewidth]{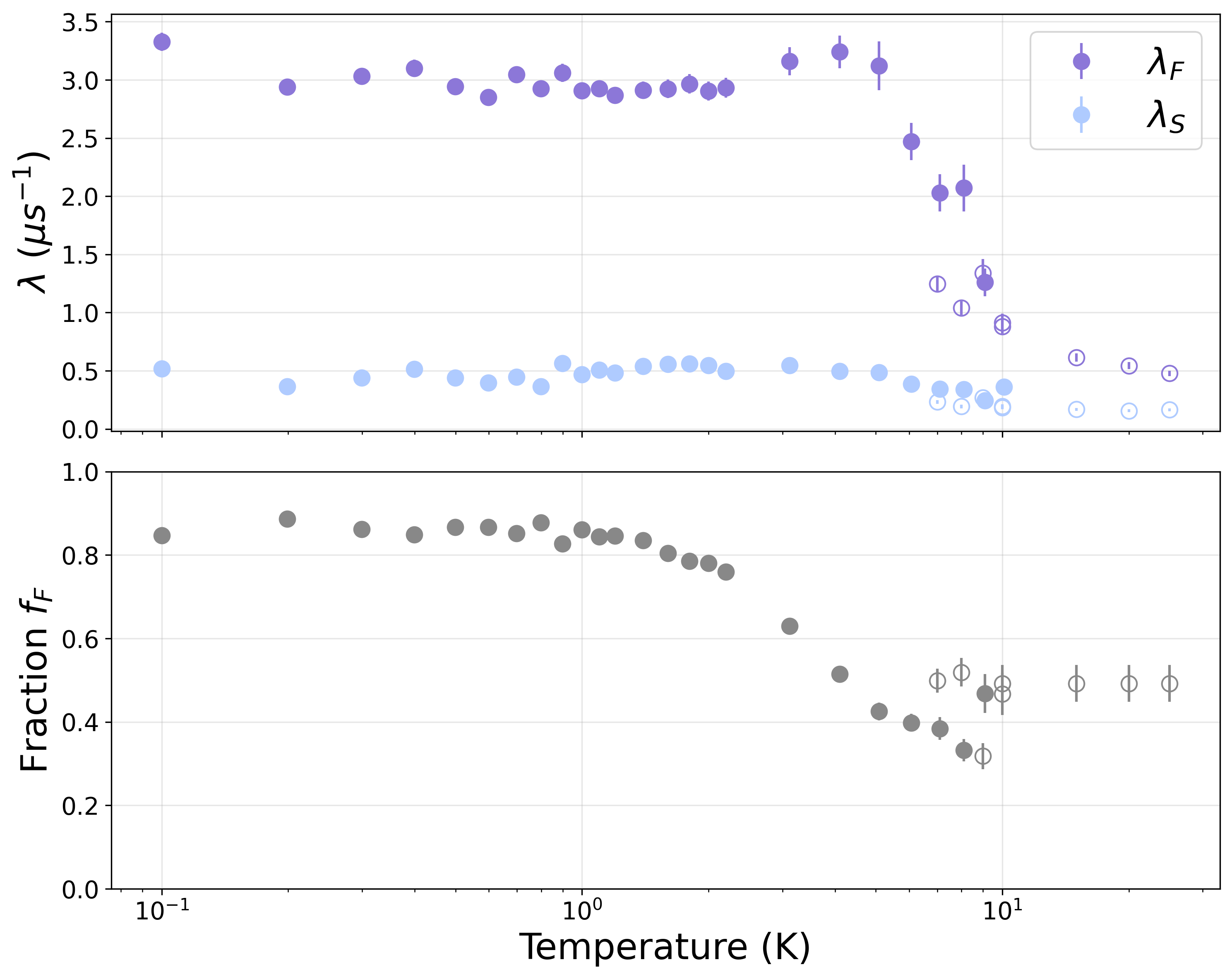}
    \caption{ZF $\mu$SR showing negligible temperature dependence between 0.1~K and 1.4~K. 
     (Top): fits to the FLAME and GPS data (see main text). (Bottom): extracted parameters $\lambda_{\mathrm{F}}$, $\lambda_{\mathrm{S}}$, and $f_{\mathrm{F}}$. 
     Solid (open) circles denote FLAME (GPS) results.}
    \label{fig:2partsfit}
\end{figure}

ZF-$\mu$SR spectra were analyzed using two different models depending on temperature. At low temperatures, the depolarization is well described by a sum of two exponential relaxation components, as shown in Fig.~\ref{fig:2partsfit}, indicating the presence of at least two distinct magnetic environments. 
\begin{equation}
\label{eq:2exponential}
A(t)
=
A_0 \left(
f_F \, e^{-\lambda_F t}
+
(1 - f_F)\, e^{-\lambda_S t}
\right)
\end{equation}
The fitting was performed with the fraction of the fast relaxation component $f_{F}$, the relaxation rate of the fast component $\lambda_{F}$, and the slow component $\lambda_{S}$ free, A$_0$ being fixed. The absence of oscillations indicates there is no long-range magnetic order. In addition, the lack of a well-defined non-relaxing $1/3$ tail shows that the longitudinal component of the muon polarization is dynamically relaxed, implying that static internal fields, if present, fluctuate on timescales fast compared to the $\mu$SR time window.

Upon increasing temperature (T$>$10 ~K), the two relaxation rates become comparable and can no longer be reliably separated. In this regime, the fraction $f_{F}$ is constrained to be a global temperature-independent fit, providing an effective description of the high-temperature dynamics and captures the increasingly homogeneous relaxation rates. 

The temperature dependence of the fitted parameters reveals a gradual slowing down of spin dynamics upon cooling. In particular, the fraction of the fast-relaxing component increases steadily at low temperatures, and its relaxation rate grows by one order of magnitude, signaling the progressive development of magnetic correlations. This crossover coincides with the broad anomaly observed in the heat-capacity data, suggesting a magnetic crossover in this intermediate temperature range. In contrast, the slow-relaxing component exhibits only a weak temperature dependence, increasing by approximately a factor of two upon cooling. This modest evolution contrasts sharply with the order-of-magnitude enhancement observed for the fast-relaxing component. One possible microscopic interpretation is that these environments arise from partially isolated or weakly correlated spins embedded within an otherwise more strongly connected magnetic network.



At high temperatures, electronic moments fluctuate rapidly in the paramagnetic regime, and the relaxation is a combination of fast electronic fluctuations and quasi-static nuclear dipolar fields.


Overall, the ZF results show no evidence for fully static spin freezing on the $\mu$SR time scale down to 40 mK: no oscillations or non-relaxing $1/3$ tail are observed, indicating that spin fluctuations persist within the $\mu$SR time window. The data further indicate a crossover with coexistence of at least two distinct spin environments. 

In general, muon depolarization reflects a distribution of fluctuating local magnetic fields characterized by a field strength $ \Delta $ and a fluctuating rate $ \nu $. In the fast-fluctuation limit, the relaxation rate follows the Redfield expression \cite{amato2024introduction}: 

\begin{equation}
\label{eq:redfield}
   \lambda =2 \Delta ^{2}/ \nu 
\end{equation}

Under the assumption that $ \Delta $ does not vary significantly with temperature, the observed increase of $ \lambda $ by a factor of 10 from 10 K to 0.1 K would correspond to a comparable reduction in the fluctuation rate $ \nu $. In contrast, at fixed temperature and for equal fluctuation rates, an increase of $\lambda$ by a factor of 7 would imply an increase in $\Delta$ by a factor of $\sqrt{7} \approx 2.7$. 

ZF measurements alone do not allow one to disentangle whether the enhanced low-temperature relaxation originates from a broadening of the field distribution, a slowing down of spin dynamics, or a combination of both. To resolve this ambiguity, longitudinal-field (LF) measurements were performed.

\subsubsection{Longitudinal-Field}

LF measurements allow one to distinguish between static and dynamic contributions to the relaxation. In the case of a purely static field distribution, complete decoupling is expected once the applied field exceeds several times the characteristic internal field scale $ \Delta $. In contrast, for dynamic fields, the relaxation rate depends on both $ \Delta $ and the fluctuation rate $ \nu $, and full decoupling may not be achieved even at high longitudinal fields. LF measurements therefore provide a means to resolve the ambiguity between broad static field distributions and slow dynamics inferred from ZF data.

\begin{figure}[]
    \centering
    \includegraphics[width=1\linewidth]{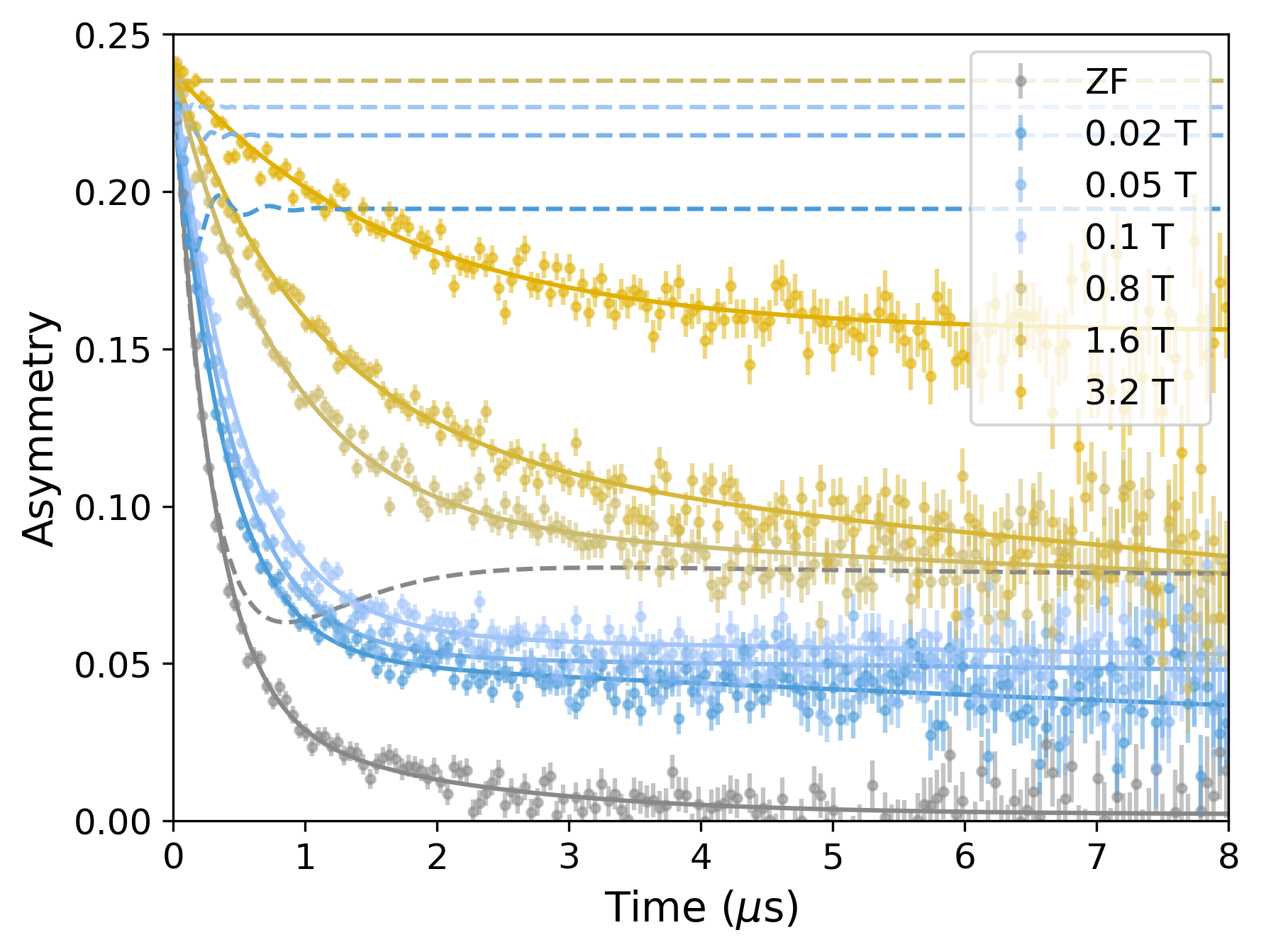}
    \includegraphics[width=1\linewidth]{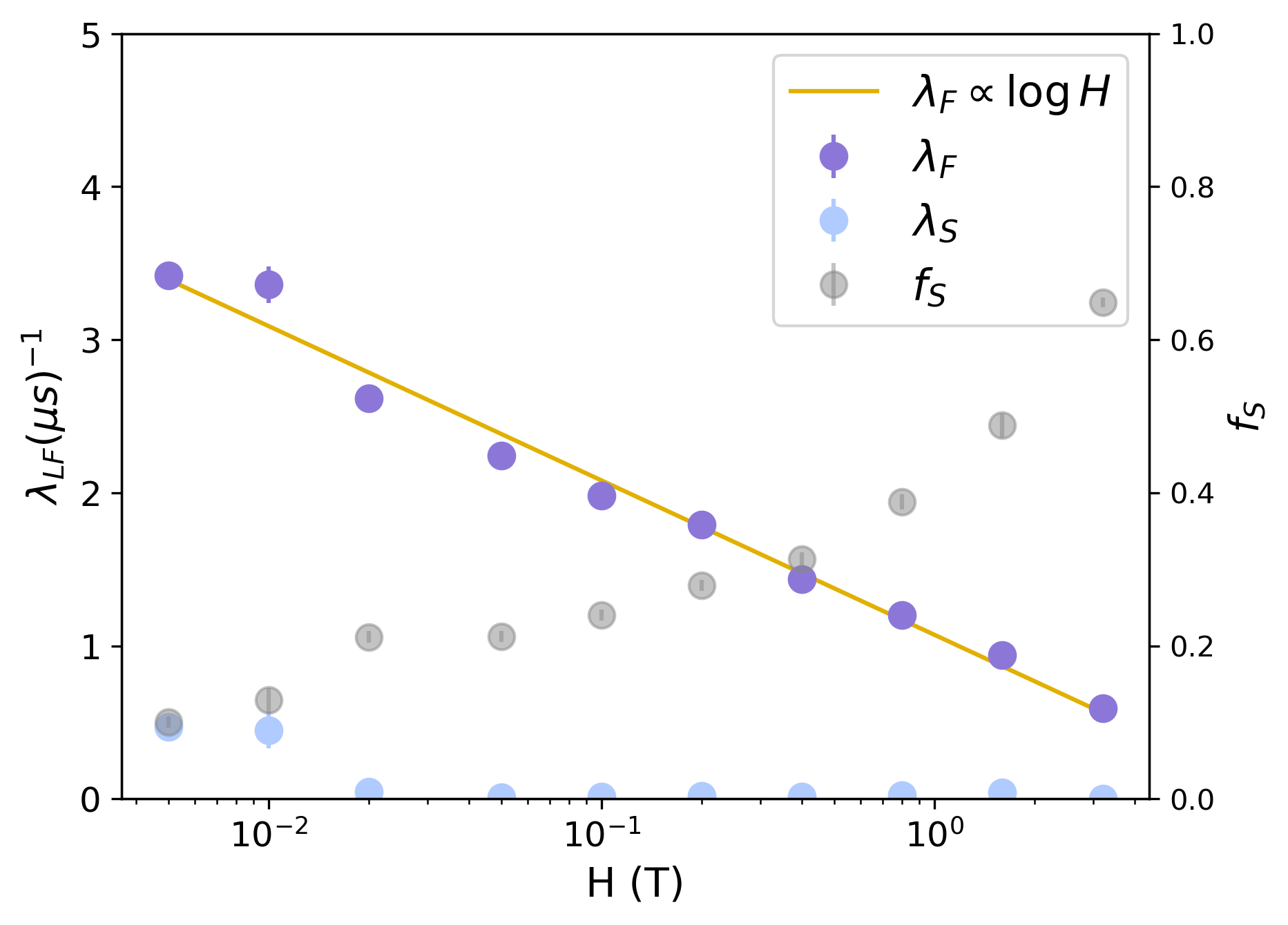}
    \caption{(Top): LF - $\mu$SR asymmetry spectra at $T = 0.1$~K, fitted with a two-exponential relaxation function (solid lines, see Eq.~\ref{eq:2exponential}). The zero-field data are additionally fitted with a static Kubo--Toyabe function, and the corresponding static relaxation is simulated for various longitudinal fields (dashed lines). The comparison demonstrates a clear deviation from a purely static ground state. (Bottom): Field dependence of the extracted relaxation rates $\lambda_F$ and $\lambda_S$, together with the fractional weight $f_S$ . The solid line is a logarithmic guide to the field dependence of $\lambda_F$.}
    \label{fig:0p1KLF}
\end{figure}

Longitudinal-field measurements were performed at 0.1 K up to 3.2 T as shown in Fig. \ref{fig:0p1KLF}. With increasing field, the muon is progressively decoupled from the internal magnetic field. However, even at 3.2 T complete decoupling is not achieved, consistent with persistent dynamics of the magnetic moments in \gBCNO{}. The asymmetry is fitted with the same two-exponential form as in zero field at 0.1~K in Fig. \ref{eq:2exponential}, with f$_{F}$, $\lambda_{F}$, $\lambda_{S}$ allowed to vary with field. The fitted results are shown in Fig. \ref{fig:0p1KLF}.


A slow relaxing component $\lambda_{\mathrm{S}}$, accounting for approximately 15--20\% of the total asymmetry at low fields, is rapidly suppressed by longitudinal fields of the order of 100~G. This behavior is consistent with muons sensing quasi-static nuclear dipolar fields (field width $\sim 10$~Oe), indicating that at these stopping sites the electronic moments fluctuate sufficiently rapidly to be in the motional-narrowing regime \cite{amato2024introduction} on the $\mu$SR time scale.

With increasing field, the corresponding non- or weakly relaxing fraction grows significantly, reflecting the progressive decoupling of these quasi-static contributions, while a substantial fraction of the signal remains dynamically relaxed over the entire field range, characterized by $\lambda_{\mathrm{F}}$.




The remaining asymmetry exhibits a gradual reduction of the relaxation rate over several decades of applied field. The field dependence of $\lambda_{F}$ does not show a well-defined decoupling scale characteristic of a single Redfield channel (see Appendix~\ref{app:musr}). Instead, $\lambda_{F}(H)$ decreases approximately linearly with $\log_{10} H$ over the measured field range, without saturation up to the highest applied field. Such a deviation from a single Redfield description indicates the absence of a well-defined characteristic fluctuation scale and is commonly interpreted in terms of a broad distribution of correlation times \cite{keren1994generalization,dunsiger2006magnetic}.

Comparison with the static Kubo–Toyabe prediction further confirms the absence of completely frozen spin: for static fields, complete decoupling is expected at longitudinal fields exceeding $\sim 10\Delta$, in clear contrast to the persistent coupled relaxation observed experimentally. The absence of complete decoupling up to 3.2 T is inconsistent with a purely static internal-field distribution and indicates persistent dynamics, indicating that the enhanced ZF relaxation originates predominantly from slowing down of magnetic moment fluctuations, rather than from a purely static, broad field distribution. This substantially reduces the ambiguity inherent in the ZF analysis and supports a dynamic origin of the relaxation.

In summary, the ZF and LF $\mu$SR results demonstrate that the magnetism in this system remains mostly dynamic down to 0.1~K. The two-exponential model captures the coexistence of at least two dynamically distinct magnetic environments, one more strongly relaxing and one more weakly relaxing. The relative spectral weights evolve continuously on cooling, with the fraction f$_{F}$ associate with the strongly relaxing component increasing at low temperature. This behavior is consistent with a gradual crossover of correlated dynamic magnetism rather than a transition into a fully frozen state.

\begin{figure*}[]
    \centering
    \includegraphics[width=\linewidth]{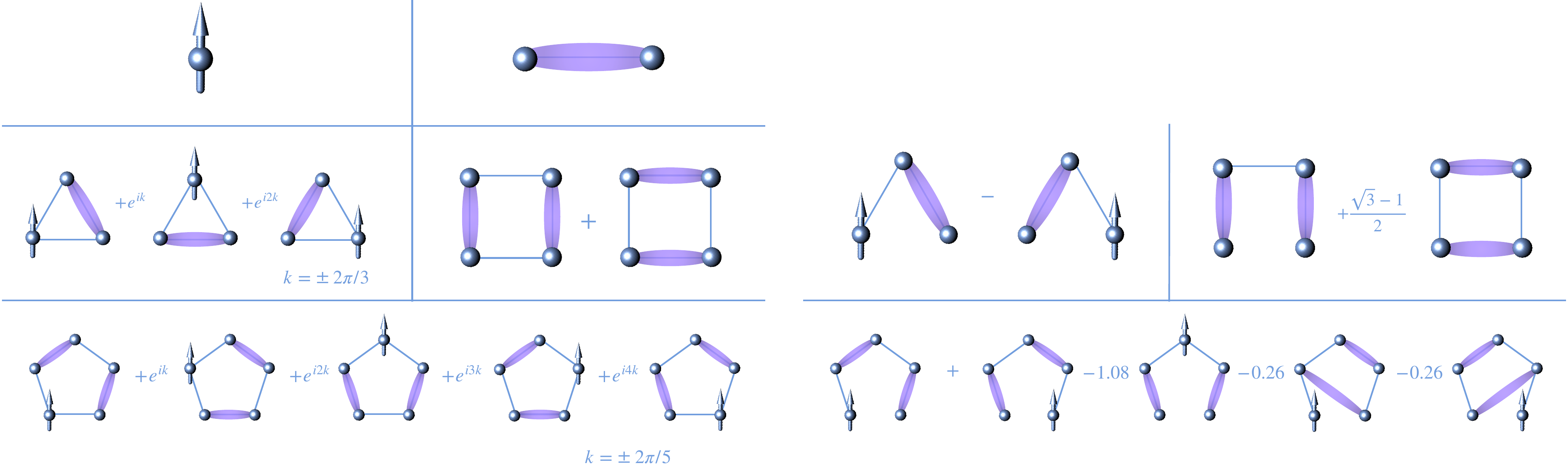}
    \caption{Ground state configurations of finite size clusters of different sizes with periodic (left) and open (right) boundary conditions. The clusters form resonating nearest-neighbour singlet (purple) coverings, and odd clusters host an uncompensated spin-1/2 degree of freedom acting as a delocalized orphan spin.}
    \label{fig:Shalf_finiteCluster}
\end{figure*}

\section{Theoretical modeling}

\subsection{$S=1/2$ Heisenberg chains}

To gain microscopic insight into the magnetic correlations in diluted cubic \gBCNO{}, we first analyze the cluster distribution. Monte Carlo simulations at $p=1/3$ yield a broad cluster‑size distribution dominated by small finite clusters (monomers, dimers, trimers, ...), but containing also some infinite, well‑connected networks, as expected above the cubic percolation threshold (see Fig.~\ref{fig:MC_cubic}). The smallest clusters are chain‑like, and assuming dominant antiferromagnetic nearest‑neighbour exchange $J_1$, these provide a controlled setting to understand how quantum singlet formation and uncompensated spins emerge on diluted networks. For $S=1/2$ Heisenberg chains with antiferromagnetic $J_1$, monomers behave as free spins and dimers form singlet ground states.  Longer chain‑like clusters realize resonating valence‑bond‑type states: even‑length clusters have total spin $S_{\rm tot} = 0$, whereas odd‑length clusters necessarily host an uncompensated spin‑1/2 degree of freedom that cannot be paired into a singlet. In this sense, odd clusters act as delocalized orphan spins embedded in a singlet background. We show the ground states for chain clusters up to five sites in Fig.~\ref{fig:Shalf_finiteCluster}, with both periodic (left panel) and open (right panel) boundaries.

With periodic boundaries, the trimer ground state is fourfold degenerate. Notably, each nearest-neighbour singlet covering of the triangle is individually an eigenstate, so the degeneracy reflects the frustration of the odd ring rather than a resonance between dimer configurations. The periodic tetramer, by contrast, is unfrustrated: its unique singlet ground state is a true resonance between the two nearest-neighbour dimer coverings of the square. For the periodic pentamer, the ground state is again a fourfold-degenerate doublet, with true resonance between the nearest-neighbour dimer coverings. For all periodic chains up to length five, the ground state can be written entirely in terms of nearest-neighbour singlets; longer-range singlets first enter for periodic chains of six or more sites. In the infinite limit, which is solved exactly by the Bethe ansatz~\cite{Bethe1931,Hulthen1938}, the periodic chain realizes a resonating valence bond state in which singlet pairs of all lengths contribute.

The clusters found in the disordered cubic lattice are perhaps more likely realized as open-boundary chains. In that case, the Lieb–Mattis theorem~\cite{lieb1962theory} guarantees the same ground-state spin quantum numbers — $S_\mathrm{tot} = 0$ for even-length and $S_\mathrm{tot} = 1/2$ for odd-length chains — and we find that the ground states of the clusters still resonate between different dimer coverings, albeit with amplitudes modified by the broken translation invariance. 

The open trimer is qualitatively different from its periodic counterpart: its ground state is a true resonating superposition of two dimer states, whereas the periodic trimer ground states can each be written as a single dimer covering. Notably, because the periodic boundary bond is absent, other dimer coverings that were purely nearest-neighbour in the ring now necessarily involve longer-range singlets in the open chain. This is already the case for the tetramer: one of the two singlet coverings pairs the end sites (a third-nearest-neighbour singlet), so the open-boundary tetramer ground state inherently contains a long-range singlet component — in contrast to the periodic tetramer, where both coverings are purely nearest-neighbour. For the open pentamer, next-nearest-neighbour singlets provide a further small correction beyond the nearest-neighbour dimer basis. In the infinite limit, the open boundaries also induce a pronounced alternation of strong and weak bond correlations that decays algebraically into the bulk~\cite{Eggert1992, Tsai2000}.

The formation of resonating valence bond states on chain clusters — with uncompensated, delocalized spin-1/2 degrees of freedom for odd-length clusters — indicates that fluctuating effective moments can naturally emerge from the interplay between structural disorder and quantum singlet formation. This provides a microscopic basis for the coexistence of weakly correlated ``orphan-like'' spins and strongly correlated singlet clusters inferred experimentally from magnetization~\footnote{If the odd clusters are to behave as true orphan spins, this would lead to a higher density of free spins than found by magnetization. In practice, the theoretical value would be modified by additional interactions, such as second-nearest neighbour couplings.} as well as for the fast spin dynamics revealed by NSE and $\mu$SR measurements.

\subsection{Minimal Heisenberg model and heat capacity}

To connect with the thermodynamic measurements, we performed exact-diagonalization (ED) simulations for a diluted spin-1/2 Heisenberg model on the simple-cubic lattice, with the Hamiltonian
\begin{equation}\label{eq:HeisenbergHamiltonian}
H = J_1 \sum_{\langle ij\rangle} \mathbf{S}_i\cdot\mathbf{S}_j
    + J_2 \sum_{\langle\langle ij\rangle\rangle} \mathbf{S}_i\cdot\mathbf{S}_j,
\end{equation}
where $J_1$ and $J_2$ denote the nearest- and next-nearest-neighbour exchange interactions, respectively, $\langle ij \rangle$ and $\langle\langle ij \rangle\rangle$ denote sums over nearest- and next-nearest-neighbour pairs on the simple-cubic lattice;  and the vector $\mathbf{S}_i$ represents a spin-1/2 on site $i$. While an antiferromagnetic $J_2$ introduces competing interactions on the simple-cubic lattice, it is treated here as a minimal extension of the nearest-neighbour model. Magnetic dilution corresponding to $p=1/3$ was implemented by randomly removing lattice sites on finite clusters, and physical observables were averaged over 100 disorder realizations. The ED calculations were carried out with periodic boundary conditions on system sizes up to 21 spins (4$\times$4$\times$4 unit cells). Full diagonalization (for thermal properties) was restricted to a maximum of 16 spins (4$\times$4$\times$3 unit cells).

\begin{figure}
    \centering
    \includegraphics[width=\linewidth]{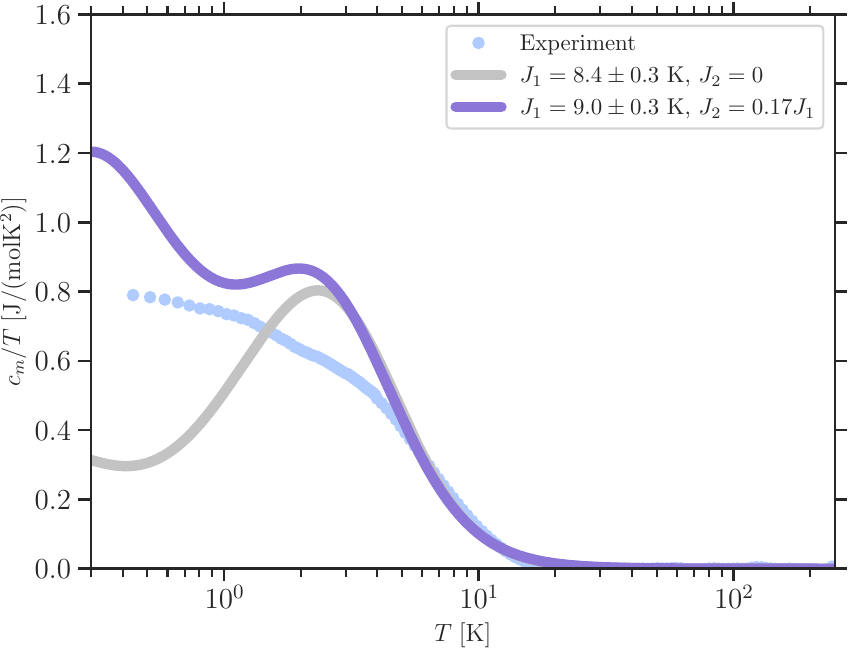}
    \caption{Numerical computation of the magnetic specific heat using ED for a 4$\times$4$\times$3 system (16 spins) for $J_2/J_1=0.17$ (purple). The value of $J_1 = 9.0~\pm ~0.3$ K is obtained from fitting with experimental data (blue) in the intermediate temperature range $T \in [3.5,36]$ K. For comparison, we also include the numerically obtained magnetic specific heat for the pure $J_1$ model (grey).}
    \label{fig:specificheat_theoretical}
\end{figure}

The calculated magnetic heat capacity $C_m(T)$ shown in Fig.~\ref{fig:specificheat_theoretical} provides important constraints on the exchange parameters. We find that the pure $J_1$ Heisenberg model fails to reproduce the broad experimental crossover in $C_m(T)$, whereas inclusion of a finite antiferromagnetic $J_2$ improves agreement with the broad thermodynamic crossover~\footnote{Note that including a small antiferromagnetic $J_2$ not necessarily breaks down the ``fluctuating spins'' picture discussed in the cluster analysis, as odd-sized clusters still will have a spin which cannot be paired into a singlet.}.

We find that including a modest antiferromagnetic 
$J_2$ ($J_2/J_1 \approx 0.15 - 0.25$) improves the qualitative agreement with the broad heat-capacity crossover below 5~K. As the further-neighbour interactions are relatively small, they can be viewed as additional weak interactions within and between the clusters.
Figure~\ref{fig:specificheat_theoretical} shows the calculated magnetic specific heat for $J_2/J_1 = 0.17$. Fitting the numerical results to the experimental data in the range $T \in [3.5,36]$~K yields only a rough estimate of the overall energy scale, $J_1 \approx 9.0$~K, with an uncertainty of about $0.3$~K. This fit should be regarded as setting the approximate exchange scale rather than providing a precise parameter determination.

We restrict the fit to the intermediate temperature regime, as the low-temperature regime is more prone
to finite-size effects: small clusters exhibit a gap-induced suppression of $C_m/T$ as $T \to 0$, but
increasing cluster size shifts this suppression to lower temperatures.  Since the ED calculations are limited to a maximum of $N = 16$ spins, clusters larger than this size are absent from the model. Such intermediate-sized clusters, which are present in the real material and contribute to the low-temperature heat capacity, are not captured by the present simulations. This omission provides an alternative explanation for the discrepancy between the pure $J_1$ model and the experimental data: the missing larger clusters may broaden the thermodynamic crossover and shift spectral weight to lower temperatures in a manner similar to the effect of introducing $J_2$, making it
difficult to uniquely determine whether $J_2$ or missing cluster sizes — or both — are responsible
for the improved agreement. Consequently, the ED results are expected to reliably capture the
thermodynamic behavior only in the intermediate temperature window (above the lowest few Kelvin),
where the broad crossover is observed, and the extracted exchange parameters should be regarded
as effective energy scales rather than uniquely determined microscopic constants~\footnote{Note that for the disordered system, the exchange couplings are likely to be dependent on the local environment, i.e. be different for each lattice site.}. We therefore infer that a disordered network of quantum spin clusters captures key aspects of the absence of a sharp thermodynamic transition and the presence of extended crossover behavior, without invoking long-range magnetic order. 


\subsection{Limitations of the minimal model}

The ED calculations reproduce the broad thermodynamic crossover, and the disorder-averaged static structure factor of the minimal $J_1$–$J_2$ model does reproduce broadened peaks (see Appendix~\ref{app:numerical_structurefactor}). However, the ED static structure factor is peaked near $(\tfrac{1}{2},\tfrac{1}{2},\tfrac{1}{2})$, as opposed to the experimentally observed peak near $(\tfrac{1}{4},\tfrac{1}{4},\tfrac{1}{4})$. This indicates that the experimentally observed modulation is not a generic consequence of simple $J_1$–$J_2$ competition alone. 

While finite-size effects in the present (quantum) numerics may contribute to the discrepancy, the absence of this modulation in the minimal model points to the relevance of additional microscopic ingredients, such as correlated chemical disorder, spatially varying anisotropic exchange induced by local octahedral distortions, bond randomness, or longer-range interactions. Resolving the microscopic origin of the modulation therefore remains an open question.

Nevertheless, the numerical results indicate that disorder strongly suppresses conventional ordering tendencies and generates a broad distribution of local correlation scales. 
In this sense, disorder qualitatively reshapes the magnetic landscape.


\subsection{Physical implications}



Overall, the theoretical results support a picture in which finite spin-$\tfrac{1}{2}$ clusters in a percolative network provide a natural source of quantum fluctuations and broad distributions of local energy scales.
The calculations indicate that (i) further-neighbor interactions refine the thermodynamic response, but are insufficient to account for the full correlation pattern; (ii) in the pure-$J_1$ limit, small finite clusters provide a nature source of fluctuating spin dynamics; and (iii) disorder strongly suppresses conventional ordering tendencies, stabilizing an extended regime of short-range correlated dynamics.



These findings support the view that dilution, proximity to the percolation threshold, and quantum spin-$\tfrac{1}{2}$ effects capture key aspects of the experimentally observed dynamically fluctuating state in a three-dimensional lattice without strong intrinsic geometric frustration.

\section{Discussion}

The combined thermodynamic and microscopic measurements establish the absence of long-range magnetic order in \gBCNO{} down to at least 0.1~K. Instead, a broad crossover regime extending from approximately 10 K to the lowest measured temperatures is observed. Within this regime, the two-component magnetization model, together with the two-component relaxation in $\mu$SR, consistently suggests the coexistence of a small fraction of weakly correlated spins and a majority of strongly correlated spin ensembles.

Direct evidence for the onset of short-range magnetic correlations is provided by both local and reciprocal-space probes. The temperature crossover identified in thermodynamic quantities is consistent with a pronounced change in the zero-field $\mu$SR relaxation rate, signalling the development of magnetic correlations on the $\mu$SR time window. Neutron powder diffraction further reveals clear diffuse magnetic scattering at low temperatures, demonstrating the presence of spatially short-range magnetic order. The diffuse peaks are modulated at wave vectors of the form $(\tfrac{1}{4},\tfrac{1}{4},\tfrac{1}{4})$, indicating a correlation pattern not captured by the minimal cubic $J_1$–$J_2$ Heisenberg model. The corresponding real-space spin–spin correlation function decays exponentially with distance (Appendix~\ref{app:Spinvert}: Fig.~\ref{fig:SSresult}), in contrast to the algebraic decay reported in some diluted frustrated three-dimensional lattices~\cite{fancelli2025fragile}. 

Crucially, the short-range magnetic order identified by neutron diffraction remains dynamical on the time scales probed by NSE and $\mu$SR; these complementary techniques access partially overlapping time windows from picoseconds to microseconds and reveal no evidence of static magnetism down to the lowest temperatures investigated. This behaviour sharply distinguishes \gBCNO{} from canonical spin-glass systems. In typical metallic spin glasses such as AuFe or CuMn, zero-field $\mu$SR experiments reveal a rapid increase of the relaxation rate already above the glass transition temperature $T_g$ \cite{uemura1985muon}, followed by the emergence of a static $1/3$ tail well below $T_g$. None of these signatures are observed here. Moreover, longitudinal-field $\mu$SR measurements demonstrate that even in a magnetic field as large as 3.2 T, the muon spin cannot be fully decoupled from the fluctuating local moments, in contrast to static magnetism, which is easily decoupled by longitudinal fields with an order of magnitude of the internal field. A consistent conclusion follows from NSE measurements: whereas spin glasses exhibit a finite long-time plateau of $S(Q,t)$ below $T_g$ and clear relaxation already observable above $\sim 2T_g$, the NSE signal of \gBCNO{} remains centered around zero without any indication of freezing down to 2 K. Together, these observations rule out a conventional spin-glass transition above this temperature and demonstrate that the crossover regime between 10 K and 5 K originates from the growth of dynamic short-range correlations rather than glassy freezing.

To explain these findings, we propose a picture in which the magnetic lattice decomposes into three components: weakly correlated orphan spins, strongly correlated finite clusters, and a well-connected infinite spin network. This scenario is supported by magnetization data, which quantify a small orphan-spin fraction, and by $\mu$SR and susceptibility measurements, which predominantly reflect the response of strongly correlated spin ensembles. Such a decomposition is a natural consequence of percolation theory for diluted cubic lattices above the percolation threshold, where an infinite cluster with high connectivity coexists with finite clusters and isolated moments.

From a microscopic perspective, the presence of fast spin dynamics in this system is particularly intriguing. The dominant interaction is the nearest-neighbour coupling on a simple cubic lattice, which lacks geometric frustration in the conventional sense. Nevertheless, strong quantum fluctuations associated with effective spin-$1/2$ moments, combined with quenched disorder and proximity to the percolation threshold, are likely to suppress static ordering. Finite clusters of spin-$1/2$ moments are expected to form singlet-based superposition states (as illustrated in Fig.~\ref{fig:Shalf_finiteCluster}), providing an intrinsic source of fluctuating spins. The nature of the infinite network remains an open theoretical challenge, as its full treatment in the presence of disorder is numerically demanding. A direct comparison with a classical-spin analogue, such as $\gamma$–Ba$_3$MnNb$_2$O$_9$ ($S=5/2$), would therefore be highly valuable for testing whether quantum fluctuations are essential for stabilizing the observed dynamical state.

Numerical studies of diluted magnetic lattices suggest that disorder enhances sensitivity to weak competing interactions, such that even modest further-neighbour couplings may qualitatively modify short-range correlations compared to the clean limit~\cite{dash2026vacancy}. 
Within the minimal $J_1$--$J_2$ model on the cubic lattice, a small antiferromagnetic $J_2$ broadens the thermodynamic crossover via weak interactions between clusters in the cluster-based fluctuating-spin picture. However, the $J_1$–$J_2$ Heisenberg Hamiltonian does not reproduce the experimentally observed modulation near $(\tfrac{1}{4},\tfrac{1}{4},\tfrac{1}{4})$. This indicates that additional microscopic ingredients, such as correlated site disorder, random anisotropies, or longer-range interactions, may influence the equal-time correlation pattern. A detailed microscopic understanding of the origin of the $(\tfrac{1}{4},\tfrac{1}{4},\tfrac{1}{4})$ modulation therefore remains an open question. 



The behavior of \gBCNO{} can be compared with other diluted cubic magnets. In Cu-based analogues such as Sr$_3$CuTa$_2$O$_9$ \cite{sana2024possible} and Sr$_3$CuNb$_2$O$_9$ \cite{hossain2024evidence}, where the orbital moment is quenched, similarly dynamical ground states have been reported from $\mu$SR measurements. The present results suggest that the unquenched orbital component in Co$^{2+}$ does not qualitatively alter the low-temperature phenomenology relative to the Cu-based analogues. This suggests that percolation physics and spin-1/2 quantum fluctuations are the essential ingredients, rather than orbital quenching.
Although local randomness inevitably introduces a distribution of single-ion anisotropies and SOC–CEF level splittings, these effects primarily broaden the relevant energy scales, rather than inducing static magnetism. 


In summary, our results on the spatial and temporal magnetic correlations of \gBCNO{} provide evidence for a disorder-driven, dynamically fluctuating magnetic state in a three-dimensional lattice without strong intrinsic geometric frustration, consistent with the important role of dilution, proximity to the percolation threshold, and spin-$1/2$ quantum fluctuations. Although this state shares key phenomenological features with QSLs---including the absence of static order and persistent fast spin dynamics---it is distinct from geometrically frustrated QSLs in that it arises primarily from quenched disorder and percolation physics, while further-neighbor interactions play only a secondary role in shaping the thermodynamic response. Future experiments with improved access to the relevant dynamical time scales, including NSE techniques with extended fast-time windows~\cite{jochum2022extending} and comprehensive longitudinal-field $\mu$SR studies~\cite{pratt2023studying}, will be crucial for distinguishing such disorder-driven dynamical states from genuine QSLs.



\section{Acknowledgements}

The authors acknowledge the Core Lab Quantum Materials at Helmholtz-Zentrum Berlin for sample synthesis and heat-capacity, magnetization, and magnetic-susceptibility measurements. This work is partially based on experiments performed at the Swiss Muon Source SµS, Paul Scherrer Institute, Villigen, Switzerland. We acknowledge the support of
the HLD at HZDR, member of the European Magnetic Field Laboratory (EMFL). We thank Diamond Light Source for access to beamline I11 under experiment cy37073. Neutron diffraction and neutron spin echo data are available at the Institut Laue-Langevin under DOIs 10.5291/ILL-DATA.5-23-818 and 10.5291/ILL-DATA.4-01-1852, respectively. FJ.X. acknowledges helpful discussions with Yuqing Ge on the $\mu$SR analysis. C.G. acknowledges funding from the European Union’s Horizon Europe research and innovation programme under the Marie Skłodowska-Curie Grant Agreement No. 101126636. We acknowledge the helpful discussion with Miguel Carvalho, Liu Hao Tjeng, and Sahana Roßler.

\appendix



\section{X-ray and neutron diffraction results}\label{sec:supplement}

To ascertain the presence of a single phase within the prepared samples, powder x-ray diffraction was measured at room temperature on a Bruker D8 diffractometer equipped with a Cu tube (40 kV, 30 mA) and a monochromator with wavelength $\lambda_{K\alpha_1}$ = 1.54059 \AA. The structure of the samples was determined by synchrotron X-ray diffraction (SXRD) and neutron powder diffraction (NPD). High resolution two-dimensional XRD patterns of both phase samples were obtained with transmission geometry on the I11 beamline at the Diamond Light Source (UK) with the MAC detector. The energy of the x-ray beam was 15 KeV and beamsize is 0.2 x 0.2 mm$^{2}$. Measurement were done at room temperature. The sample used in synchrotron-XRD was diluted with crushed boron silica and sealed in a 0.3 mm diameter capilary (boron silica). Neutron diffraction pattern of \gBCNO{} was obtained on the High-resolution two-axis diffractometer D2B at the Institut Laue-Langevin (France) with the wavelength of 1.594 \AA. Measurement were done at 1.5~K and room temperature. 

\begin{figure}[H]
    \centering
 \includegraphics[width=1\linewidth]{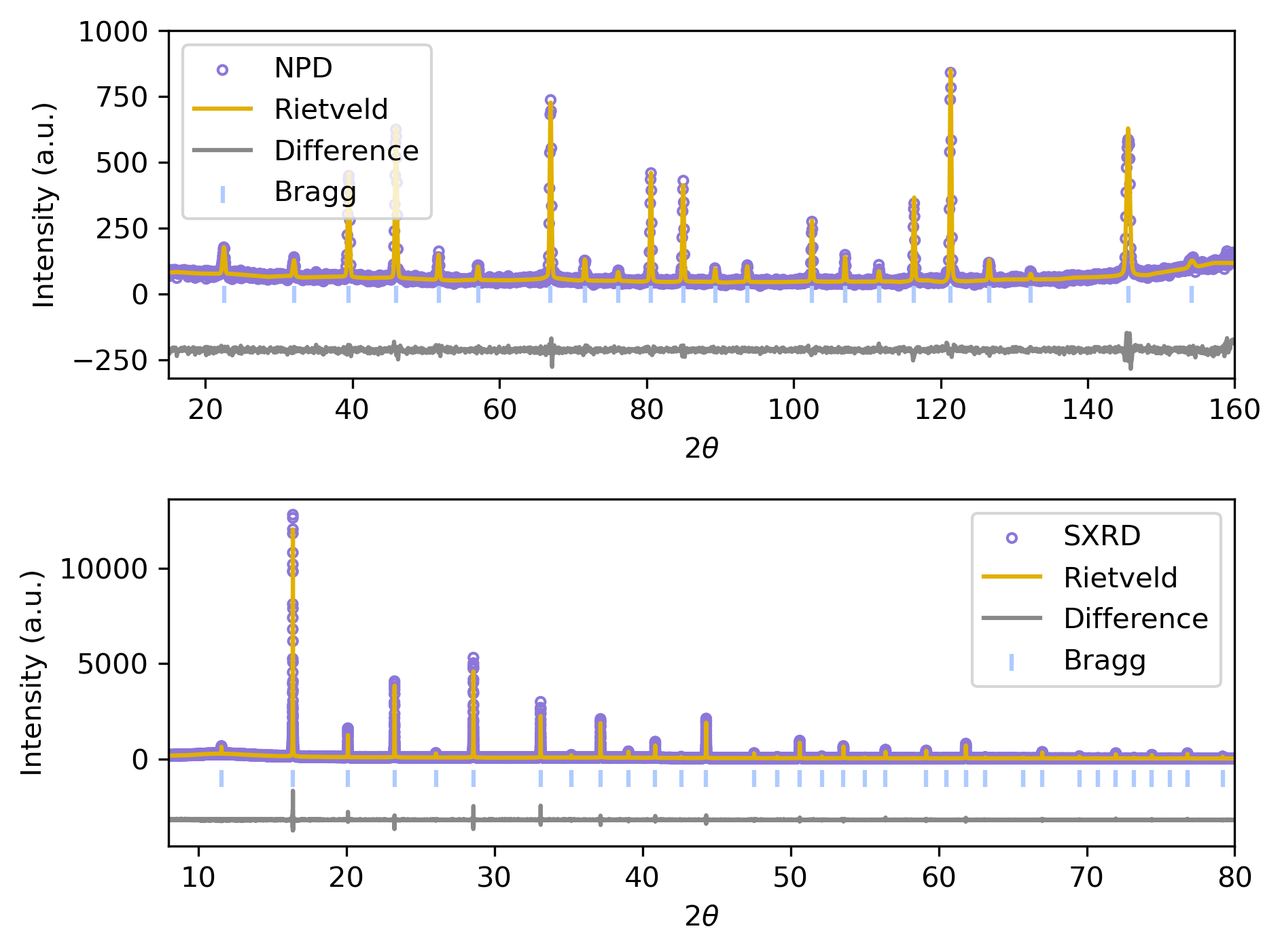}    
    \caption{(Top): Neutron powder diffraction at room temperature (purple circles), Rietveld refinement (solid yellow line), difference between the observed and calculated pattern (solid purple line) and the Bragg positions (blue vertical lines) are plotted. Space group Pm-3m, with RF = 3.159, Bragg R-factor: 4.456, $\chi^{2}$= 6.53. (Bottom): Synchrotron diffraction pattern at room temperature (purple circles), Rietveld refinement (solid yellow line), difference between the observed and calculated pattern (solid purple line) and the Bragg positions (blue vertical lines) are plotted. Space group Pm-3m, with RF = 6.720, Bragg R-factor: 6.980, $\chi^{2}$= 2.38.}
    \label{fig:enter-label}
\end{figure}

\begin{table}[H]
    \centering
    \begin{tabular}{ |c|c|c|c|c|c|c| } 
           \hline
         Atom & Wyckoff & x & y & z & B & occ.\\
         \hline
         \hline
          Ba& 1a & 0 & 0 & 0 & 0.728(4) & 1\\ 
         Co& 1b & 0.5 & 0.5 & 0.5 & 0.728(4) & 1/3\\ 
         Nb& 1b & 0.5 & 0.5 & 0.5 & 0.728(4) & 2/3\\ 
         O & 3c & 0 & 0.5 & 0.5 & 0.728(4) & 1\\
         \hline
         \end{tabular}
    \caption{Atomic coordinates (x, y, z) and isotropic Debye-Waller factors (B) obtained from the rietveld refinement of the synchrotron (MAC-I11, Diamond) and neutron (D2B, ILL) powder diffraction pattern at room temperature. The refinement was performed in space group Pm-3m (No. 221) yielding lattice parameters a = b = c = 4.0859(7)  \AA.}
    \label{tab:table1}
\end{table}

\section{$\mu$SR additional data}\label{app:musr}


To further examine the longitudinal-field (LF) dependence of the relaxation rates, we attempted to parameterize the data within the framework of the Redfield model for fluctuating local magnetic fields. For a single dynamical channel characterized by a field distribution width $\Delta$ and fluctuation rate $\nu$, the LF relaxation rate is given by

\begin{equation}
\label{LFredfield}
\lambda(H)
=
\frac{2 \gamma_\mu^{2} \Delta^{2} \nu}
{\nu^{2} + \gamma_\mu^{2} H^{2}}
\end{equation}

where $\gamma_\mu$ is the muon gyromagnetic ratio. However, single-component Redfield fits do not reproduce the gradual suppression of $\lambda_F$ over the full measured field range. We therefore considered a multi-component Redfield description.

\begin{equation}
\lambda_{\mathrm{tot}}(H)
=
\sum_{i} w_i \,
\frac{2 \gamma_\mu^{2} \Delta_i^{2} \nu_i}
{\nu_i^{2} + \gamma_\mu^{2} H^{2}}
\end{equation}

Here, $w_i$ are the fractional weights ($\sum_i w_i = 1$), $\Delta_i$ is the local-field distribution width of the $i$-th channel, $\nu_i$ is the corresponding fluctuation rate, $\gamma_\mu$ is the muon gyromagnetic ratio, $H$ is the applied longitudinal field.

\begin{figure}[H]
    \centering
    \includegraphics[width=1\linewidth]{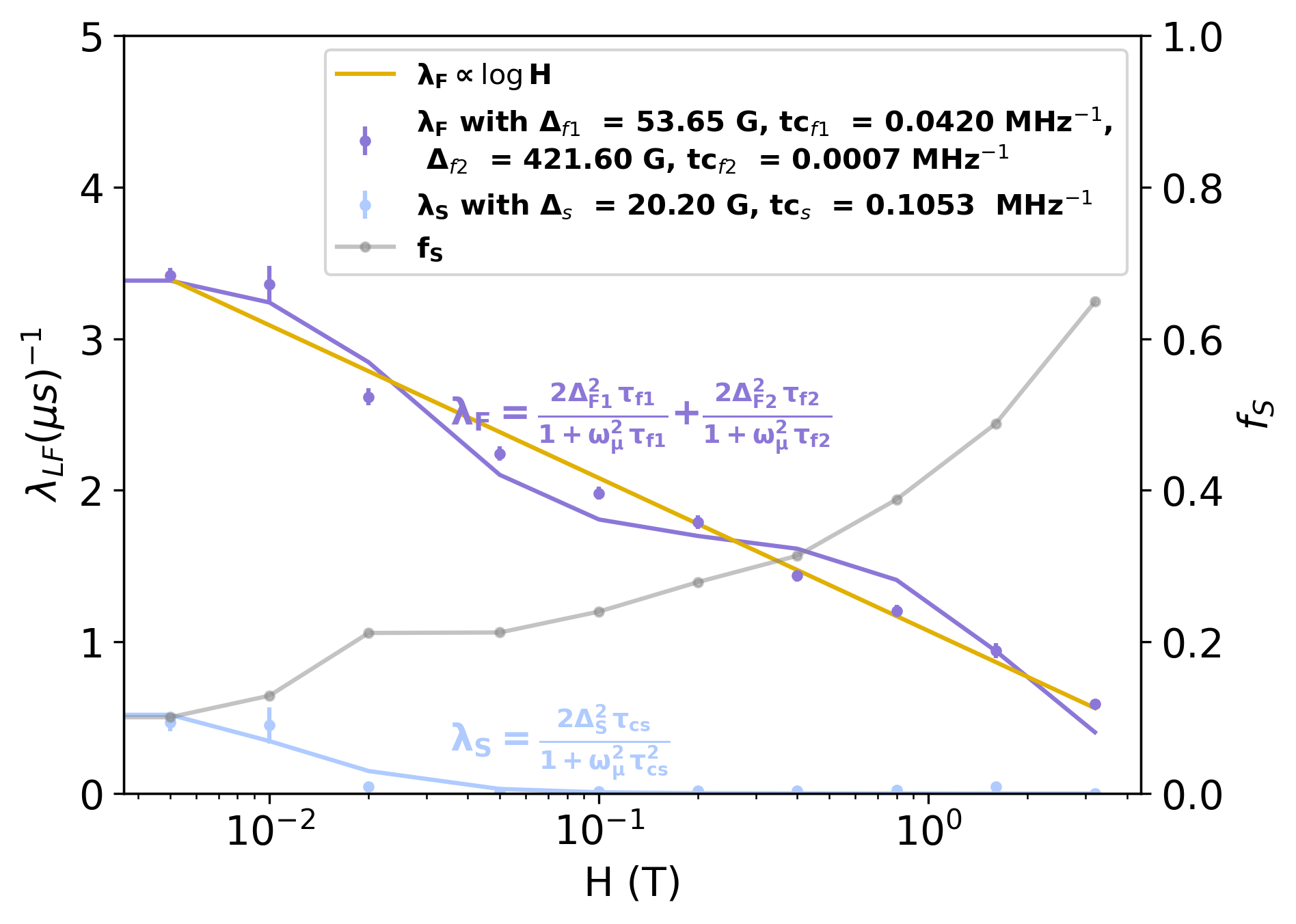}
    \caption{ Relaxation rate $\lambda_F$ and $\lambda_S$ extracted from two-exponential time-domain fits at 0.1~K, plotted as a function of longitudinal field. The solid line represents a two (for $\lambda_F$) and one (for $\lambda_S$)-component Redfield parameterization}
    \label{fig:0p1KLFstsaticKT}
\end{figure}

Among the tested models (one-, two-, and three-component Redfield forms), the two-component model provides the best and most stable representation of the data. Introducing a third component does not lead to a statistically significant improvement and instead results in strong parameter correlations. For completeness, representative parameter sets obtained from two-component Redfield fits are shown in Fig. \ref{fig:0p1KLFstsaticKT}. 

We emphasize that the parameters obtained from the multi-component Redfield fits should be regarded as phenomenological and not assigned a strict microscopic interpretation. Owing to intrinsic correlations between the fitted quantities ($\lambda_F$, $\lambda_S$, and $f_S$, see Eq. \ref{eq:2exponential}), as well as the pronounced field dependence of the fractional weight $f_S$, the extracted relaxation rates do not represent independent measures of distinct dynamical processes, but instead provide an effective description of the overall field evolution of the relaxation. Consequently, while the multi-component Redfield model captures the general trends in the data, the fitted parameters should not be over-interpreted in terms of well-defined microscopic fluctuation rates or local field distributions.









\section{Thermodynamic properties}
\subsection{Effective moment of Co$^{2+}$} \label{app:Comoment}
The analytical expression of the effective moment of Co$^{2+}$ taking account the SOC-CEF effect is developed by Lloret et al \cite{lloret2008magnetic}. 

\[
\mu_{\mathrm{eff}}^{2}(T,\lambda)
=
3\mu_B^{\,2}\,\frac{
\begin{aligned}
&\left(c_1 + c_2\,\frac{ k_B T}{\lambda}\right)
\\[4pt]
&+ \left(c_3 + c_4\,\frac{ k_B T}{\lambda}\right)
   \exp\!\left(\frac{9\lambda}{4 k_B T}\right)
\\[4pt]
&+ \left(c_5 + c_6\,\frac{ k_B T}{\lambda}\right)
   \exp\!\left(\frac{6\lambda}{k_B T}\right)
\end{aligned}
}{
1 
+ 2 \exp\!\left(\frac{9\lambda}{4 k_B T}\right)
+ 3 \exp\!\left(\frac{6\lambda}{k_B T}\right)
},
\]

where $\lambda$ is the spin-orbit coupling constant, with the following coefficients:
\[
c_i = 
\left\{
\frac{169}{36},\;
-\frac{490}{81},\;
\frac{128}{45},\;
\frac{4312}{2025},\;
\frac{63}{20},\;
\frac{98}{25}
\right\},
\qquad
\lambda < 0.
\]

\subsection{High-field magnetization}\label{app:Highfield}

\begin{figure}[H]
    \centering
 \includegraphics[width=1\linewidth]{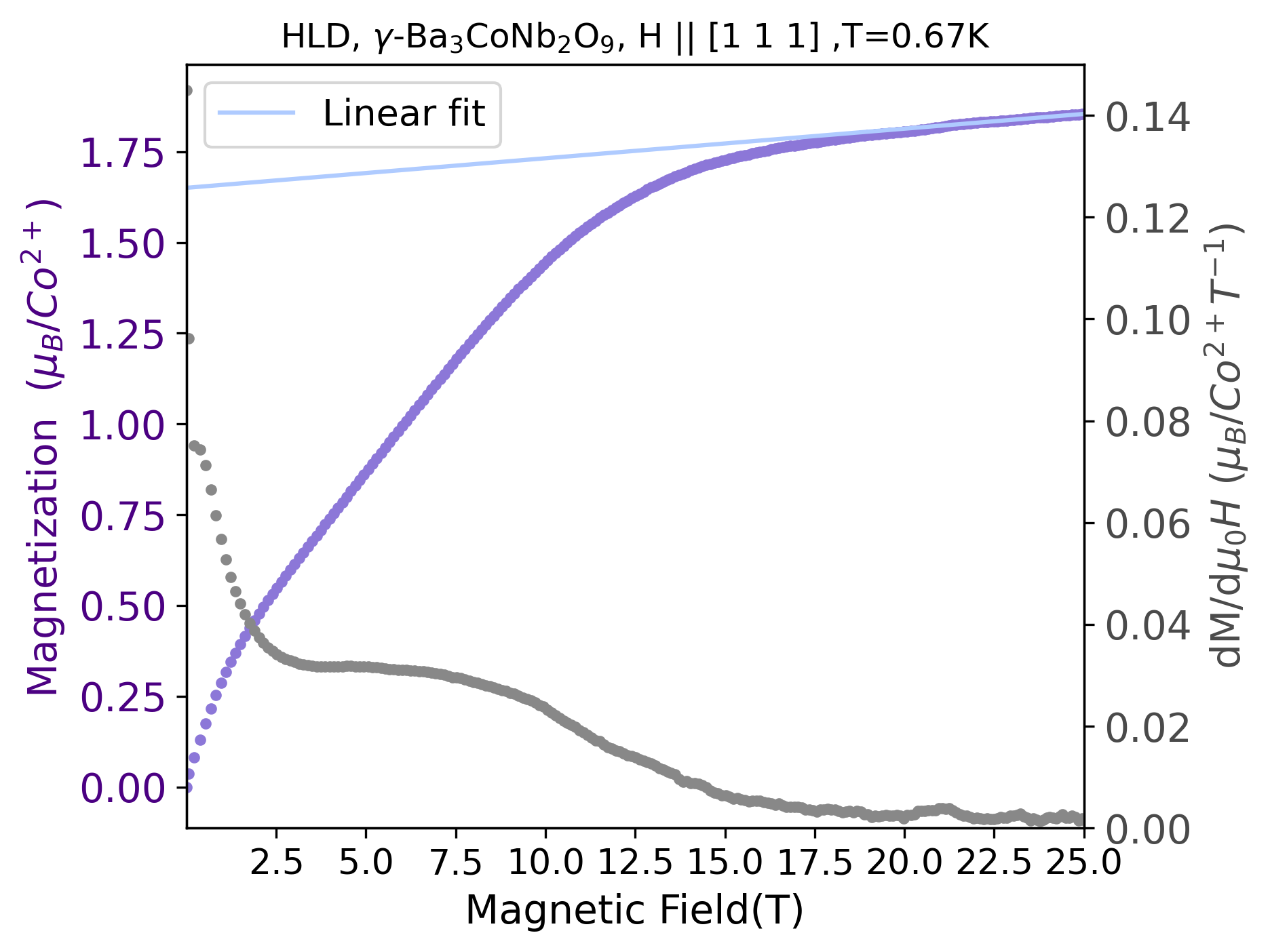}    
    \caption{High-field magnetization. The linear increase above 16 T are fitted with a linear form $M(H) = aH + b$. Here, $a$ represents the field-independent Van Vleck susceptibility, and $b$ corresponds to the saturated moment. The fitted parameters are $a=0.081 \pm 0.003$~($\mu_B$/Co$^{2+}$T$^{-1}$), $b=1.650 \pm 0.001$~($\mu_B$/Co$^{2+}$).}
    \label{fig:enter-label}
\end{figure}

\section{Spinvert additional result}\label{app:Spinvert}

\begin{figure}[H]
    \centering
    \includegraphics[width=1\linewidth]{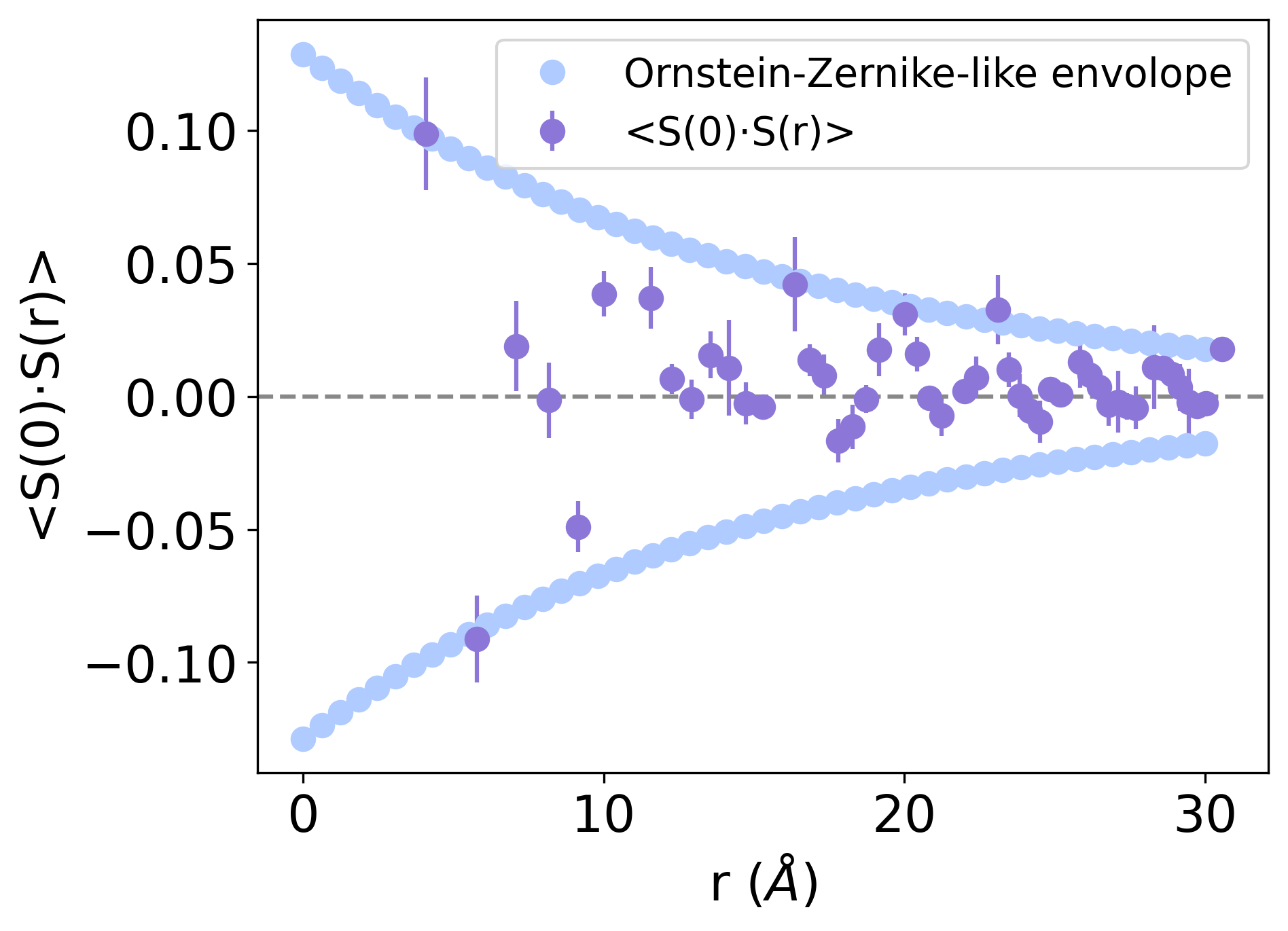}
    \caption{Spin-spin correlation function $\langle S(0)·S(r)\rangle$ values for increasing interatomic distances at 1.5 K.}
    \label{fig:SSresult}
\end{figure}

\section{Numerical static structure factors}\label{app:numerical_structurefactor}

We compute the ground-state static structure factor of the $J_1$-$J_2$ Heisenberg Hamiltonian in Eq.~\eqref{eq:HeisenbergHamiltonian} for both quantum and classical spins. In the classical limit, the clean system exhibits a sharp Bragg peak at $J_2/J_1$-dependent ordering wave vectors, which are readily obtained by minimizing the Fourier-transformed exchange coupling
\begin{equation}\label{eq:Jq}
\begin{split}
    J_{\vec{q}} = &\; J_1 (\cos q_x + \cos q_y + \cos q_z )\\ 
    + 2&J_2 (\cos q_x \cos q_y  + \cos q_y \cos q_z  + \cos q_z \cos q_x ).
\end{split}
\end{equation}
The result (in reciprocal lattice units) is:
\begin{itemize}
    \item $J_2/J_1 < 1/4$: $(\tfrac{1}{2},\tfrac{1}{2},\tfrac{1}{2})$ (N\'eel order),
    \item $J_2/J_1 = 1/4$: $(h,\tfrac{1}{2},\tfrac{1}{2})$ for arbitrary $h$ (degenerate lines),
    \item $J_2/J_1 > 1/4$: $(0,\tfrac{1}{2},\tfrac{1}{2})$ (collinear antiferromagnetic order),
\end{itemize}
and symmetry-equivalent wave vectors. At the critical ratio $J_2/J_1 = 1/4$, the energy $J_{\vec{q}}$ is independent of $h$, giving rise to a one-dimensional manifold of degenerate ground states. The clean classical structure factor thus shows a sharp peak at the ordering wave vector of height $N$, where $N$ is the number of sites.

We then move on to investigate how the result of the clean classical model is amended by both disorder and quantum fluctuations. We restrict the study to ground-state properties and use iterative energy minimization (aligning each spin with its local field) for the classical case, and ED via the Lanczos method \cite{lanczos1950iteration} for the quantum case. Since the experimental structure factor shows a peak near the wave vector $(\tfrac{1}{4},\tfrac{1}{4},\tfrac{1}{4})$, we require this momentum to be commensurate with the finite-sized lattice. The smallest supercell resolving the desired wave vector is the 4$\times$4$\times$4 system. With a site occupancy $p=1/3$, this system contains $64 \times (1-p) \approx 21$ magnetic sites. For the quantum spin-$\tfrac{1}{2}$ model, this is the largest system amenable to ED resolving the wave vector $(\tfrac{1}{4},\tfrac{1}{4},\tfrac{1}{4})$. We average over 100 independent random disorder configurations.

\begin{figure}
    \centering
    \includegraphics[width=\linewidth]{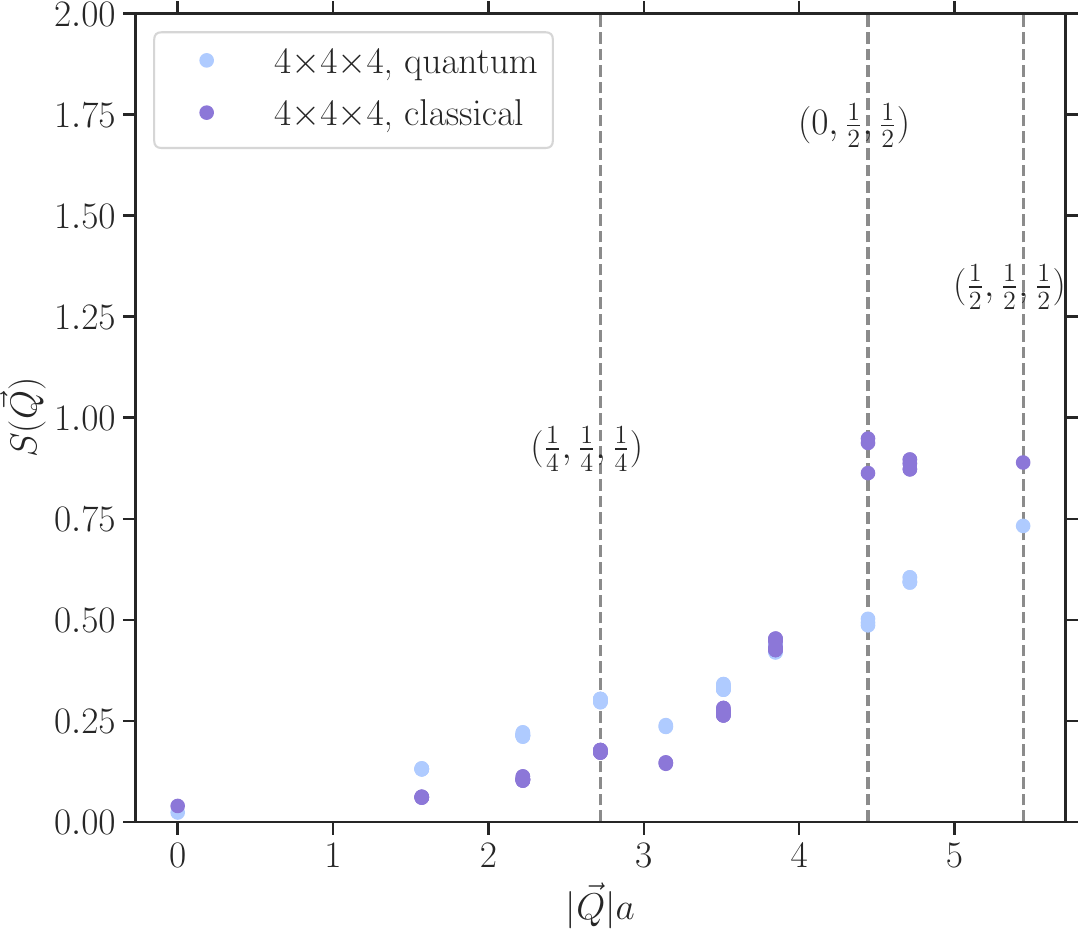}
    \caption{Ground-state structure factor of the quantum ($S=\tfrac{1}{2}$) and classical disordered systems on a 4$\times$4$\times$4 system (21 sites) with $J_2/J_1 = 0.17$.}
    \label{fig:quantum_vs_classical}
\end{figure}

\begin{figure}
    \centering
    \includegraphics[width=\linewidth]{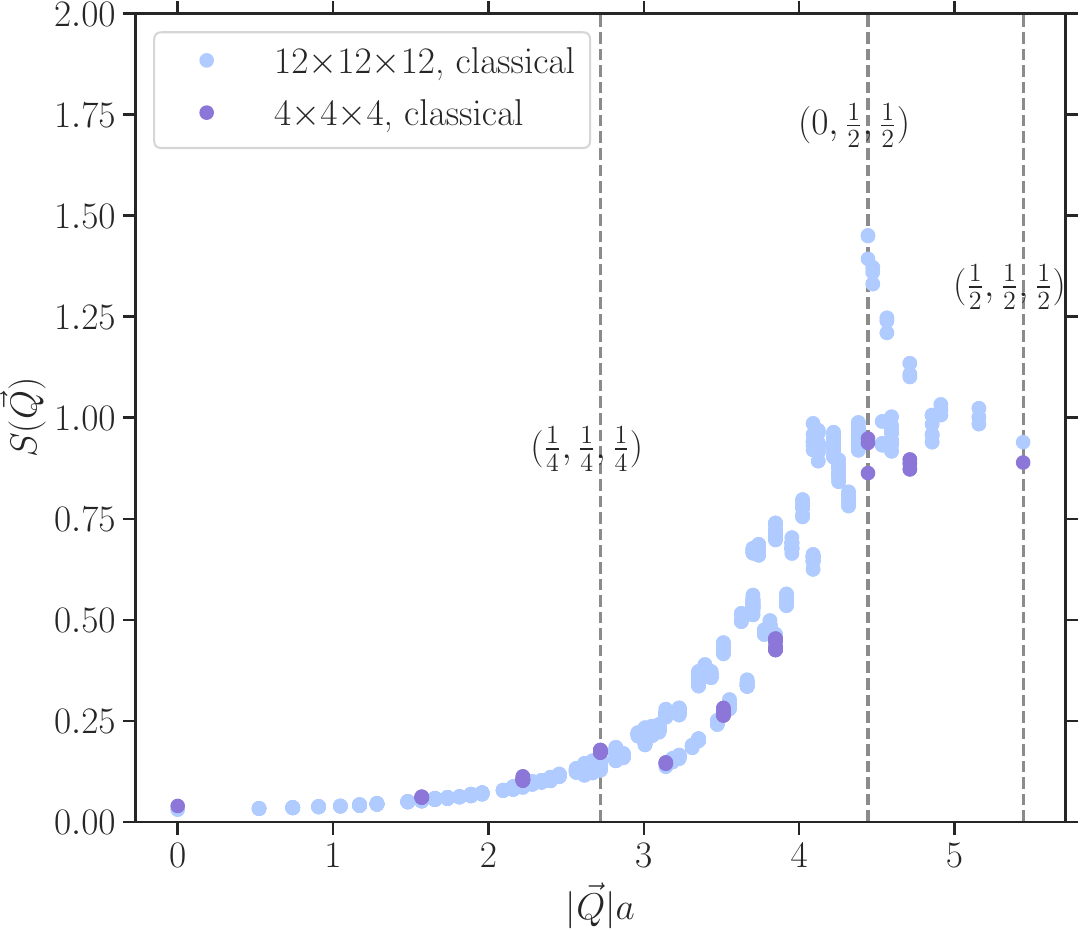}
    \caption{Ground-state structure factor of two disordered classical systems on 4$\times$4$\times$4 (21 sites) and 12$\times$12$\times$12 (576 sites) supercells, both with $J_2/J_1 = 0.17$.}
    \label{fig:finite_size}
\end{figure}

In Fig.~\ref{fig:quantum_vs_classical}, we show the obtained ground state structure factors for the disordered classical and quantum systems on the 4$\times$4$\times$4 supercell with $J_2/J_1 = 0.17$.
In contrast to the long-range ordered classical ground states, both the classical and quantum disordered systems exhibit broad peak features in the structure factor. Disorder alone can thus explain broadening of the structure factor. However, the maxima of the structure factors are not at $(\tfrac{1}{4},\tfrac{1}{4},\tfrac{1}{4})$, but rather at $(\tfrac{1}{2},\tfrac{1}{2},\tfrac{1}{2})$ and $(0,\tfrac{1}{2},\tfrac{1}{2})$ for the quantum and classical models, respectively.

Figure~\ref{fig:quantum_vs_classical} does however show a local maximum at $(\tfrac{1}{4},\tfrac{1}{4},\tfrac{1}{4})$. To assess finite-size effects, we compare the classical ground-state structure factor of the 4$\times$4$\times$4 system (21 sites) with that of a 12$\times$12$\times$12 system (576 sites) in Fig.~\ref{fig:finite_size}. The hump feature present near $(\tfrac{1}{4},\tfrac{1}{4},\tfrac{1}{4})$ in the smaller system becomes significantly less prominent in the larger system, indicating that it is a finite-size artifact in the classical case. While this suggests that the corresponding feature in the quantum model may also be a finite-size effect, this conclusion does not follow necessarily, as quantum fluctuations can stabilize correlations that are absent classically.

\newpage

\nocite{*}

\bibliography{myreferences}

\end{document}